\pdfoutput=1

\documentclass[11pt]{article}

\usepackage[final]{acl}

\usepackage{times}
\usepackage{latexsym}

\usepackage[T1]{fontenc}

\usepackage[utf8]{inputenc}

\usepackage{microtype}

\usepackage{inconsolata}

\usepackage{graphicx}

\usepackage[utf8]{inputenc} 
\usepackage[T1]{fontenc}    
\usepackage{hyperref}       
\usepackage{url}            
\usepackage{booktabs}       
\usepackage{amsfonts}       
\usepackage{nicefrac}       
\usepackage{microtype}      
\usepackage{xcolor}         
\usepackage{array}
\usepackage{amssymb}
\usepackage{enumitem}
\usepackage{graphicx}
\usepackage{subcaption}
\usepackage{threeparttable}
\usepackage{pifont}
\usepackage{amsmath}
\usepackage{multirow}
\newcommand{\xmark}{\ding{55}}

\usepackage{bbding}

\definecolor{red}{RGB}{238, 68, 51}
\definecolor{blue}{RGB}{70, 177, 225}
\definecolor{yellow}{RGB}{255, 192, 0}
\definecolor{purple}{RGB}{216, 110, 204}
\definecolor{brown}{RGB}{127, 36, 28}
\definecolor{green}{RGB}{71, 172, 20}
\definecolor{orange}{RGB}{194,153,107}

\usepackage{listings}
\usepackage{framed}

\title{AMEX: Android Multi-annotation Expo Dataset for Mobile GUI Agents}

\author{
 \textbf{Yuxiang Chai\textsuperscript{1}\textsuperscript{*}},
 \textbf{Siyuan Huang\textsuperscript{2}\textsuperscript{*}},
 \textbf{Yazhe Niu\textsuperscript{1}},
 \textbf{Han Xiao\textsuperscript{1}},
\\
 \textbf{Liang Liu\textsuperscript{3}},
 \textbf{Guozhi Wang\textsuperscript{3}},
 \textbf{Dingyu Zhang\textsuperscript{1}},
 \textbf{Shuai Ren \textsuperscript{3}},
 \textbf{Hongsheng Li \textsuperscript{1}\textsuperscript{\textdagger}},
\\
\\
 \textsuperscript{1}MMLab @ CUHK,
 \textsuperscript{2}SJTU,
 \textsuperscript{3}vivo AI Lab,
\\
 \textsuperscript{*}Equal contribution,
 \textsuperscript{\textdagger}Corresponding author: hsli@ee.cuhk.edu.hk
}

\begin{document}
\maketitle
\begin{abstract}
AI agents have drawn increasing attention mostly on their ability to perceive environments, understand tasks, and autonomously achieve goals. To advance research on AI agents in mobile scenarios, we introduce the Android Multi-annotation EXpo (AMEX), a comprehensive, large-scale dataset designed for generalist mobile GUI-control agents which are capable of completing tasks by directly interacting with the graphical user interface (GUI) on mobile devices. AMEX comprises over 104K high-resolution screenshots from popular mobile applications, which are annotated at multiple levels. Unlike existing GUI-related datasets, e.g., Rico~\cite{Deka:2017:Rico}, \textsc{AitW}~\cite{rawles2024androidinthewild}, etc., AMEX includes three levels of annotations: GUI interactive element grounding, GUI screen and element functionality descriptions, and complex natural language instructions with stepwise GUI-action chains. We develop this dataset from a more instructive and detailed perspective, complementing the general settings of existing datasets. Additionally, we finetune a baseline model SPHINX Agent and illustrate the effectiveness of AMEX.
\end{abstract}

\section{Introduction}

In recent years, AI-powered virtual assistants, such as Siri, Bixby, and Xiao AI, have evolved as key tools for facilitating interactions between users and mobile devices. These assistants have demonstrated to be effective in managing routine tasks such as setting alarms, performing web searches, and reporting weather conditions. However, their functionality is often restricted to system-built applications or third-party apps supported by application programming interfaces (APIs). This limitation highlights a significant gap between the capabilities of current AI assistants and the diverse, open-ended ways humans interact with mobile devices.

Unlike AI assistants, human users can accomplish tasks on mobile devices by relying solely on visual information from the screen. Humans intuitively interpret graphical user interfaces (GUIs), analyze page layouts, and infer the functionalities of interactive elements to complete tasks in various apps. Inspired by this human ability, researchers are exploring new paradigms for mobile interaction that leverage vision and natural language processing. One such paradigm involves the development of Mobile GUI-Control Agents, or GUI Agents, which aim to directly interact with screen elements based on visual and textual inputs, such as screenshots and natural language instructions. These agents hold the potential to transcend the limitations of traditional API-based assistants, enabling more flexible and universal interactions with any app or mobile interface.

Existing GUI agents show promise but face significant challenges in real-world applications. Their effectiveness is possibly limited by a lack of understanding of GUI layouts and the functionalities of interactive elements through the observation of wrong predictions. These limitations largely arise from the absence of comprehensive datasets that adequately reflect the complexity and diversity of mobile GUI environments. While several instructional datasets~\cite{burns2021mobile, sunkara-etal-2022-towards, gubbi-venkatesh-etal-2024-ugif, rawles2024androidinthewild} have been introduced to address this issue, they struggle with limitations such as inaccurate annotations, insufficient task diversity, and a lack of representativeness for general mobile usage. Similarly, other efforts~\cite{Deka:2017:Rico, li-etal-2020-mapping, mud, bunian2021vins} that focus on GUI element annotation are constrained by limited dataset scale, outdated app versions, and traditional annotation styles that do not provide sufficient information for GUI agents.

\begin{figure*}[t]
    \centering
    \includegraphics[width=0.9\linewidth]{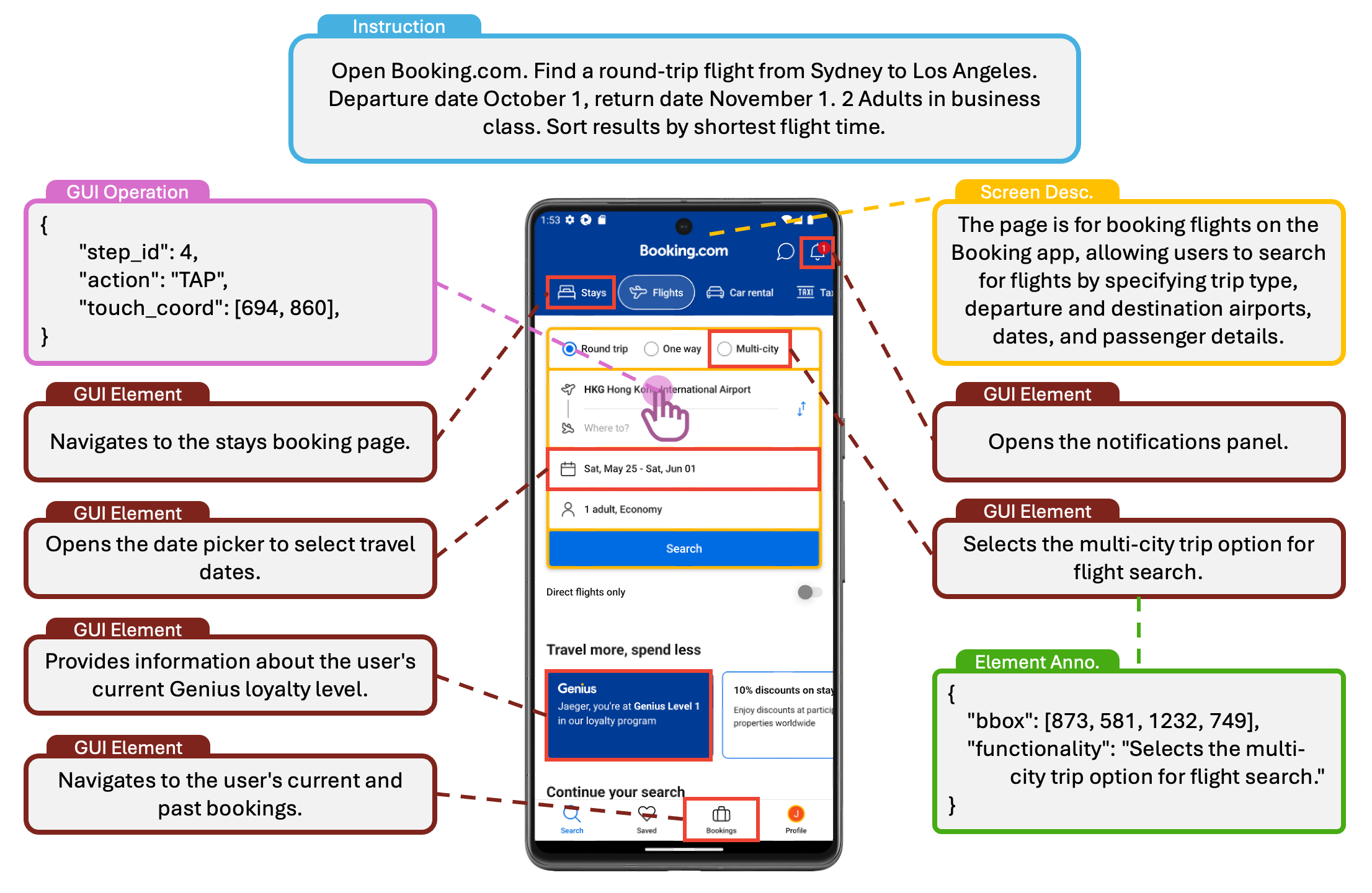}
    \caption{An example of a screenshot-instruction (\textcolor{blue}{Blue} tab) pair illustrating the multi-level annotation of AMEX. \textcolor{red}{Red} boxes + \textcolor{brown}{Brown} tabs: selected GUI interactive elements and their corresponding functionalities. \textcolor{green}{Green} tab: the detailed annotation of the element. \textcolor{yellow}{Yellow} tab: the description of the entire screenshot. \textcolor{purple}{Purple} hand icon + tab: the current action and the annotation.}
    \label{fig:main-contribution}
\end{figure*}

Human users approach GUI-based tasks by integrating multiple cognitive processes. They interpret the overall layout of the page, identify interactive elements and assess their functionality, such as recognizing that a bell icon likely leads to notifications and a magnifier icon likely triggers a search function. Users then decompose natural language instructions into actionable steps, navigating across multiple screens to achieve their goals. To design effective GUI agents, it is essential to replicate these human cognitive processes, which requires datasets that comprehensively capture the structure, functionality, and interaction patterns of mobile GUIs.

To address this need, we introduce a new dataset, the \textbf{A}ndroid \textbf{M}ulti-annotation \textbf{EX}po (AMEX), specifically designed to advance the development of GUI agents by providing a multi-level understanding of mobile GUIs. AMEX includes three levels of annotations: (i) GUI interactive element grounding, (ii) GUI screen and element functionality descriptions, and (iii) instructions with GUI-action chains. The dataset comprises over 104K high-resolution screenshots, 21K screen descriptions with 300K element-wise functionalities, and approximately 3,000 unique complex instructions, with an average of 12.8 steps (see Figure~\ref{fig:main-contribution}, Table~\ref{tab:dataset-comparison}, and Table~\ref{tab:dataset-stats}). To ensure annotation precision and quality, all annotations, including bounding boxes, screen and element functionalities, and instructions with GUI-action chains, are verified by human annotators trained with detailed guidelines.

Our contributions can be summarized as follows: (a) We collect and release AMEX, a multi-level dataset, providing reliable and detailed understandings of the smartphone GUI environment, which can serve as a strong supplementary dataset to boost the performance of GUI agents; (b) We train SPHINX Agents, which can serve as the baseline models for future researches on GUI agents and illustrates the effectiveness of AMEX.

\section{Related Work}

\subsection{GUI-related Datasets}
\label{sec:gui-related-datasets}

\newcolumntype{C}[1]{>{\centering\let\newline\\\arraybackslash\hspace{0pt}}m{#1}}

\begin{table*}[t]
  \centering
  \resizebox{\linewidth}{!}{%
  \begin{threeparttable}
    \caption{Comparison of mobile GUI-related datasets. \textbf{Scale}: the number of unique instructions on general third-party apps, average steps per instruction, number of screenshots and element functionality descriptions. \textbf{Diversity}: screenshot description, element labels, element functionality and the action details for stepwise operation.}
    \label{tab:dataset-comparison}
    \begin{tabular}{l C{1.1cm} C{1.1cm} C{1.3cm} C{1.1cm} C{1.4cm} C{1.6cm} C{2.0cm} C{1cm}}
        \toprule
        Dataset &
        Screen\newline Desc. &
        Screen\newline Element &
        Element\newline Func. &
        Task \&\newline Action &
        \# Screen-\newline shots &
        \# Element\newline Func. &
        \# Unique\newline General Inst. & 
        \# Avg\newline Steps \\
        \midrule

        RICO & \xmark & \checkmark & \xmark & \xmark & 72K & - & - & - \\
        RICO semantics & \xmark & \checkmark & \xmark & \xmark & 72K & - & - & - \\
        VINS & \xmark & \checkmark & \xmark & \xmark & 4K & - & - & - \\
        MUD & \xmark & \checkmark & \xmark & \xmark & 18K & - & - & - \\
        PixelHelp & \xmark & \xmark & \xmark & \checkmark & 800 & - & 187 & 4.2 \\
        UGIF & \xmark & \xmark & \xmark & \checkmark & 3.3K & - & 480 & 6.3 \\
        MoTIF & \xmark & \xmark & \xmark & \checkmark & 21K & - & 480 & 4.5  \\
        \textsc{AitW} & \xmark & \textbf{*} & \xmark & \checkmark & 510K & - & 1539 & 6.5 \\
        \textsc{AitZ} & \checkmark & \textbf{*} & $\blacksquare$ & \checkmark & 18K & 18K & 2504 & 7.5 \\
        \textsc{AndroidControl} & \xmark & \textbf{*} & \xmark & \checkmark & 99K & - & 15,283 & 4.8 \\
        \midrule
        \textbf{AMEX} & \checkmark & \checkmark & \checkmark & \checkmark & 104K & 296K & 3046 & 12.8 \\
        \bottomrule
    \end{tabular}
    \begin{tablenotes}
        \small
        \item * indicates elements are mis-annotated and the bounding boxes are not well-aligned.
        \item $\blacksquare$ indicates containing only action results of the element interacted at each step. 
    \end{tablenotes}
  \end{threeparttable}
  }
\end{table*}

\begin{table*}
  \caption{AMEX statistics}
  \label{tab:dataset-stats}
  \centering
  \resizebox{0.9\linewidth}{!}
{
  \begin{tabular}{cccccc}
    \toprule
    \# Screenshots & \# Apps & \# Interactive Elements & \# Functionalities & \# Instructions & \# Avg. Steps \\
    \midrule
    104,876 & 192 & 1,659,647 & 296,075 & 3046 & 12.8\\
    \bottomrule
  \end{tabular}
  } 
\end{table*}

Table~\ref{tab:dataset-comparison} compares several popular GUI-related datasets on Android. While some works~\cite{deng2024mind2web, liu2018reinforcement, shi2017world, wu2023webui} focus on the web or desktop platforms, on Android OS, many works~\cite{Deka:2017:Rico, li-etal-2020-mapping, mud, bunian2021vins} have focused on identifying various types of GUI elements, assigning classes to different elements. This annotation style can be defined as a traditional icon classification and detection problem. In many cases, however, the functionality of an icon depends on the context of the GUI and many elements on mobile devices are not an icon and their functionalities are also critical. Besides, most of the datasets~\cite{Deka:2017:Rico, bunian2021vins, li-etal-2020-mapping} are outdated, where the screenshot interfaces are completely different from modern apps. The most recent MUD~\cite{mud} only provides 18K screenshots, which are insufficient for understanding the layouts and elements for LLMs.

Other studies~\cite{burns2021mobile, rawles2024androidinthewild, zhang-etal-2024-android, gubbi-venkatesh-etal-2024-ugif} primarily emphasize action-observation pairs during instructional operations, but their annotations are limited and often require supplemental View Hierarchy (VH) data for each screenshot. \textsc{AitW}~\cite{rawles2024androidinthewild} provides both screen GUI element annotations and instructions, but it includes only a small portion of instructions for third-party apps, with most operations conducted on Chrome and other system-built apps. Additionally, each instruction is repeated multiple times, resulting in significant data redundancy. In response, \textsc{AitZ}~\cite{zhang-etal-2024-android} filters \textsc{AitW} thoroughly, selecting 2.5K unique instructions and episodes, and introduces the Chain-of-Action-Thought framework to annotate action results and page descriptions better. Despite this refinement, \textsc{AitZ} contains only 18K screen-action pairs. \textsc{AndroidControl}~\cite{li2024androidcontrol} is another large-scale instruction-based dataset, however, the element-wise annotations contain heavy redundancy and misaligned cases due to the lack of human verification.

\begin{figure*}
    \centering
    \includegraphics[width=0.98\linewidth]{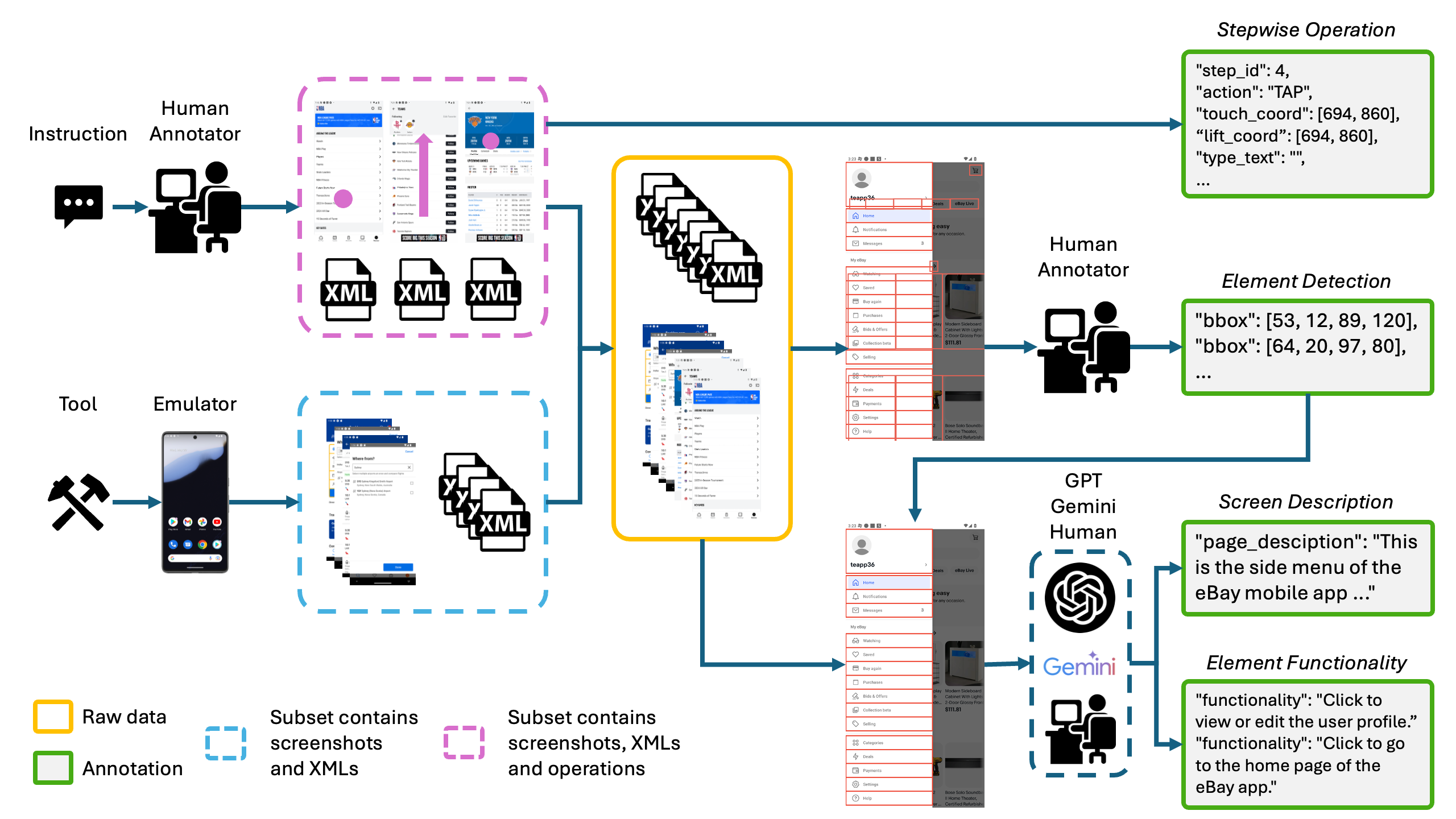}
    \caption{Overview of the data collection pipeline. The raw data is from two subsets collected by human annotators and an autonomous tool. Annotators record the GUI-action chains simultaneously while collecting the screenshots and corresponding XMLs. Then raw data is sent to annotators to filter GUI element bounding boxes, and then the boxes with corresponding raw screenshots are sent as input to GPT and Gemini to extract the GUI screen and element descriptions, which are then manually checked by humans.}
    \label{fig:pipeline}
\end{figure*}

\subsection{GUI-related Systems and Agents}

~\cite{wang2023enabling} first starts to apply Large Language Model (LLM) on GUI, but it remains on tasks only interacted on single page, which are more like question-answer tasks rather than end-to-end instructional tasks. Recent advancements have leveraged the extensive world knowledge and robust embodied capabilities of LLMs~\cite{gu2024mamba, gur-etal-2023-understanding, touvron2023llama} for task planning and reasoning.  Works~\cite{yang2023appagent, pmlr-v235-zheng24e} have utilized business-level models (e.g., GPT-4V), employing extensive prompt engineering to guide the LLM in executing complex tasks. The effectiveness of these methods requires a meticulous prompt design to achieve optimal results. Alternatively, another research line focuses on fine-tuning smaller LLMs on GUI-specific datasets to imbue them with domain-specific knowledge. For example, CogAgent~\cite{hong2024cogagent} enhances performance in GUI-related tasks by integrating a high-resolution cross-module that fuses image features from various levels. CoCo-Agent~\cite{ma2024coco} and \textsc{AndroidControl}~\cite{li2024androidcontrol}, unlike other agents taking only screenshots as input, use element layouts from accessibility trees or view hierarchy as additional input to enhance the performance. However, still many apps don't support accessibility information or only provide very little of that.

\begin{figure*}[t]
    \centering
    \begin{subfigure}[b]{0.52\linewidth}
        \centering
        \includegraphics[width=0.501\textwidth]{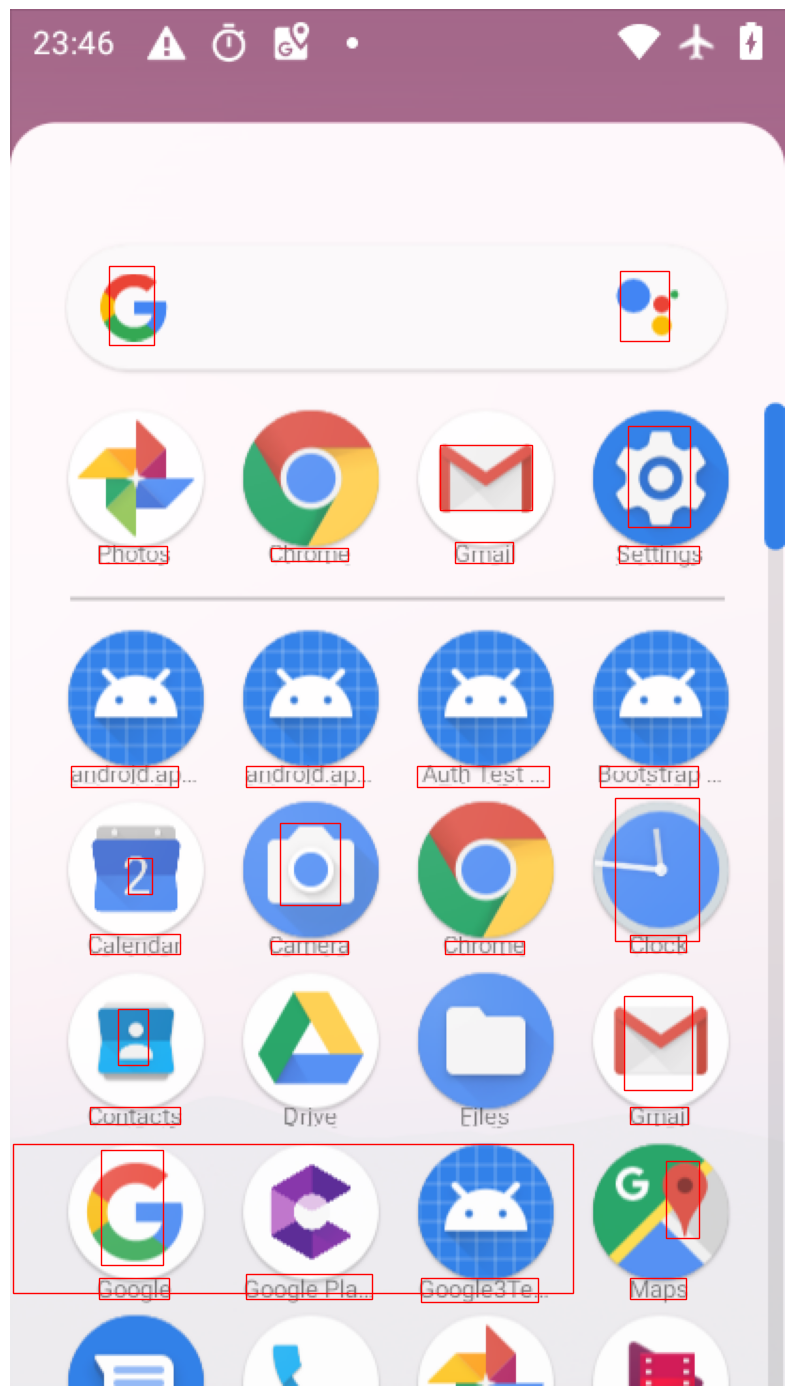}
        \includegraphics[width=0.447\textwidth]{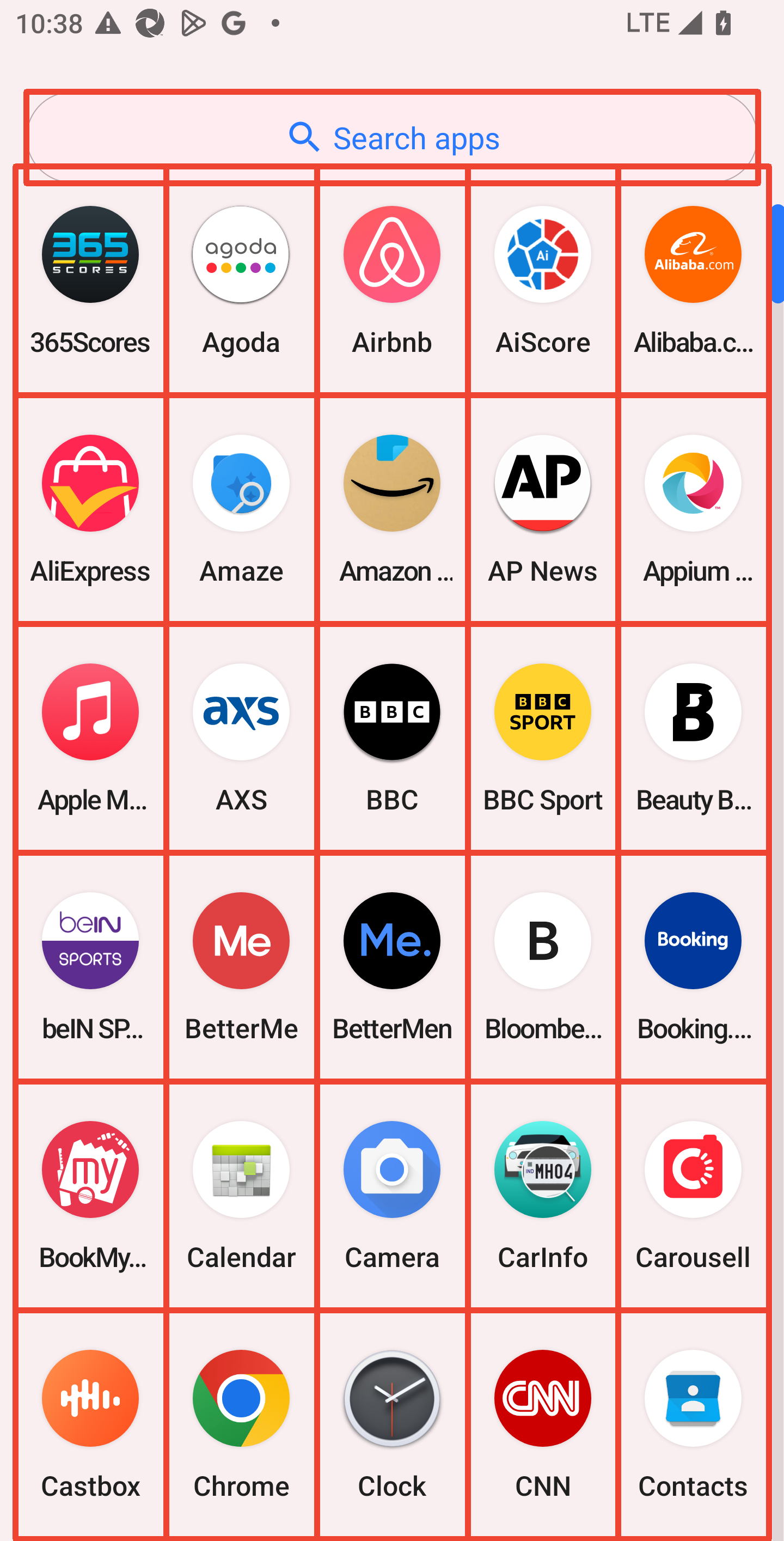}
        \caption{Element annotation on \textsc{AitW} (left) and AMEX (right).}
        \label{fig:aitw-amex}
    \end{subfigure}
    \hfill
    \begin{subfigure}[b]{0.45\linewidth}
        \centering
        \includegraphics[width=0.48\textwidth]{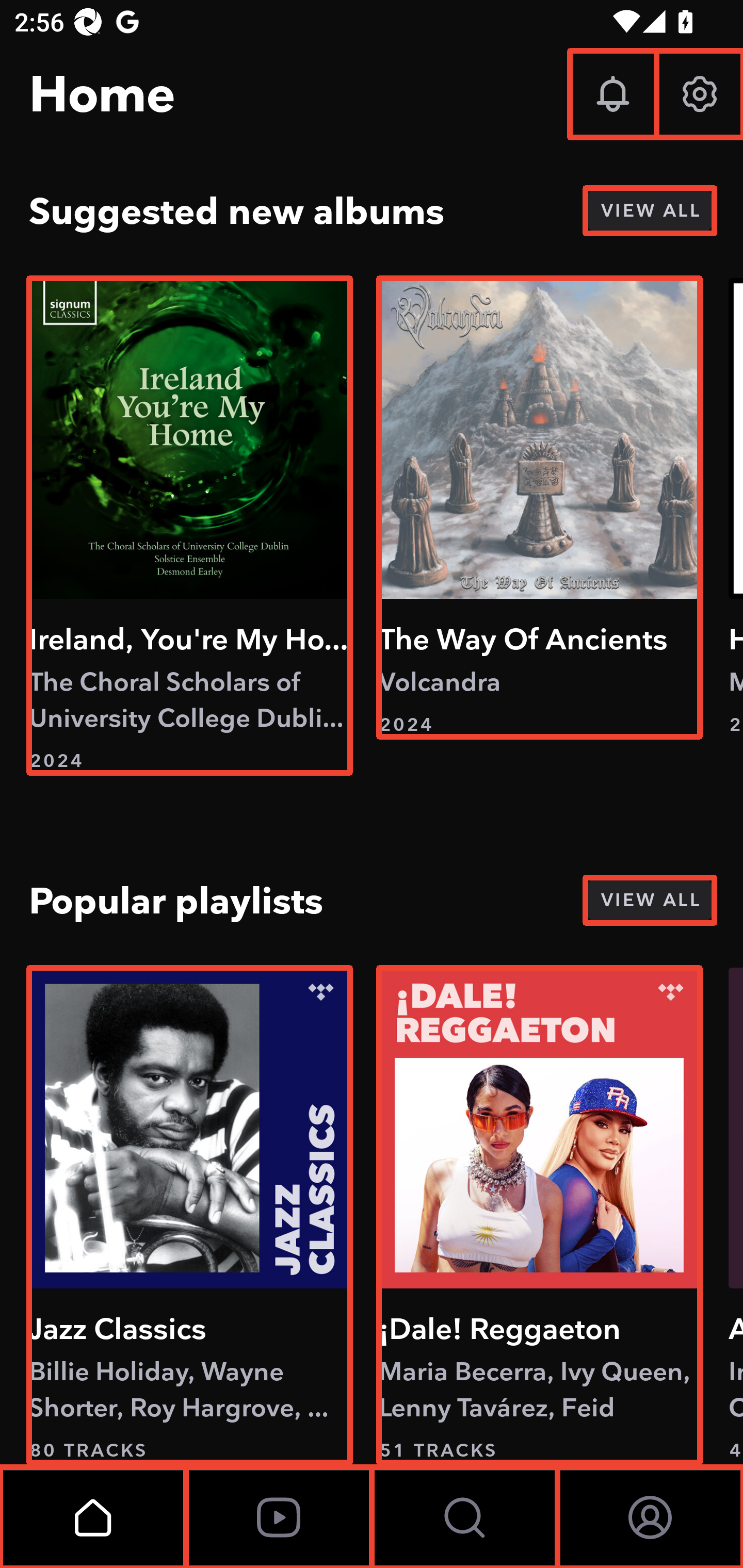}
        \includegraphics[width=0.48\textwidth]{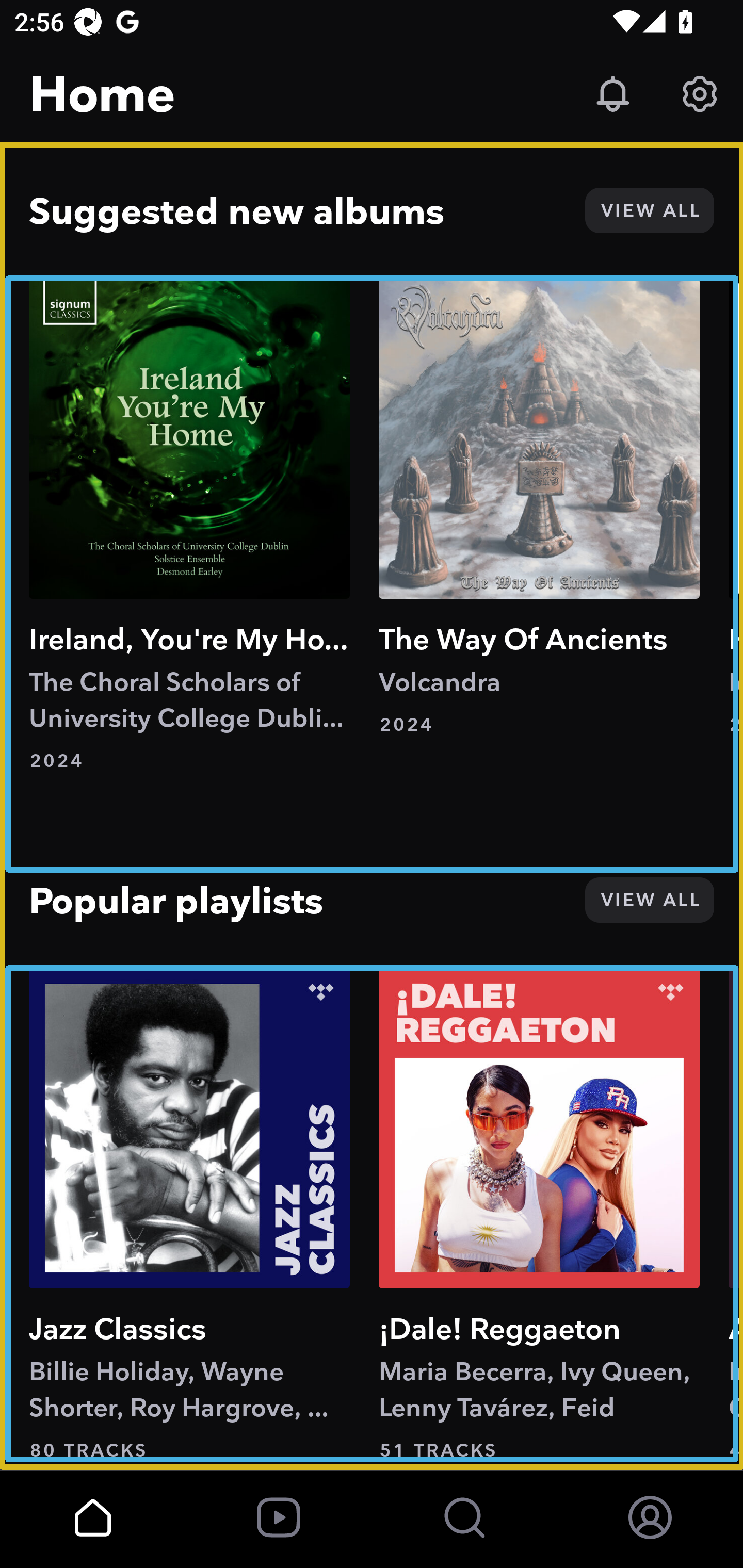}
        \caption{Demo of GUI interactive elements.}
        \label{fig:amex-elements}
    \end{subfigure}
    \caption{Demonstrations of element annotations of \textsc{AitW} and AMEX. (a) Element bounding boxes in \textsc{AitW} (left) and AMEX (right). Boxes in AMEX are well aligned but boxes in \textsc{AitW} might be misaligned and mis-annotated. (b) GUI interactive elements in AMEX. \textcolor{red}{Red} boxes: clickable elements. \textcolor{blue}{Blue} boxes: horizontally scrollable elements. \textcolor{yellow}{Yellow} box: vertically scrollable element.}
\end{figure*}

\section{Android Multi-annotation EXpo (AMEX)}
\label{sec:amex}

When receiving an instruction, a human user first analyzes the overall screen layout to form a basic understanding of the current Android environment. The user then identifies interactive elements and areas, and assesses the functionalities of those elements. Finally, the user breaks down the instruction into simple, step-by-step actions on each screen. Based on this human cognitive process, we design three levels of annotations in our proposed AMEX training set: (i) GUI interactive element grounding (see Section~\ref{sec:gui_interactive_element_grounding}), (ii) GUI screen and element descriptions (see Section~\ref{sec:element-function}), and (iii) instructions with GUI-action chains (see Section~\ref{sec:instrtions_with_gui_operations}). The overall statistics of AMEX are listed in Table~\ref{tab:dataset-stats}.

\subsection{Data Collection Pipeline}
\label{sec:data-collection}

An overview of the data collection pipeline is illustrated in Figure~\ref{fig:pipeline}. The raw data is collected through two methods: human instruction-following GUI manipulations and autonomous GUI controls. Human GUI manipulations involve recording stepwise operations for each instruction, and simultaneously storing screenshots and each screen's Extensible Markup Language (XML) data. In parallel, an autonomous script controls emulators to collect additional screenshots and their XMLs. These two subsets comprise the entire raw dataset. Then for each screenshot, initial bounding boxes of interactive elements and their in-app descriptions (if available) are parsed from the corresponding XML. Human annotators then review each screenshot to filter out all the misaligned boxes, which serve as the interactive element grounding annotations. With the in-app descriptions, GPT and Gemini generates the functionalities of the selected elements and provides descriptions for the whole screenshot. Annotators then further check the quality of the descriptions of functionalities. More details are discussed in the following sections and Appendix~\ref{app:collection-details}.

\subsection{Level I: GUI Interactive Element Grounding}
\label{sec:gui_interactive_element_grounding}

Existing datasets~\cite{Deka:2017:Rico, sunkara-etal-2022-towards, mud, bunian2021vins} typically classify elements on the screen, such as icons, texts and images, based on their types. Instead of adhering to the traditional classification paradigm, we define interactive elements more broadly as any elements that users can interact with, regardless of their specific types (see Figure~\ref{fig:aitw-amex}). Specifically, interactive elements in our dataset are only categorized into two subsets: (i) clickable elements and (ii) scrollable elements. The motivation of our annotation is mainly from two facts: (i) no existing dataset provides information on the ``scroll'' action, which is crucial especially for agents to know the horizontally scrollable area. (ii) ``click'' action is the most used action and many ``compound'' elements cannot be simply categorized. As illustrated in Figure~\ref{fig:amex-elements} red boxes, four ``compound'' elements in the middle contain both image and text and the definition of ``clickable'' is more suitable with the functionality in Level 2 for agents to understand the element.

Clickable Elements are the most common components in a screen. They typically include clickable icons, images, texts, and compounds that combine several categories. We also include certain ``typeable'' elements, such as search bars, because most typeable elements require a prior click action to enable typing.

Scrollable Elements typically occupy larger areas on the screen. In most cases, a scrollable element area contains many clickable elements and supports a pair of actions, such as ``scroll down'' and ``scroll up'' or ``scroll left'' and ``scroll right.'' Figure~\ref{fig:amex-elements} illustrates these two types of scrollable elements.

\subsection{Level II: GUI Screen and Element Functionality Descriptions}
\label{sec:element-function}

Previous works~\cite{Deka:2017:Rico, mud, bunian2021vins} on GUI elements often rely on predefined class names, such as ``image'' and ``text'', to convey the underlying meaning of each element. Rico Semantics~\cite{sunkara-etal-2022-towards} annotates the icons with slightly more detailed semantics information but the annotations are still classifications on icons. However, this classification-based method has significant limitations. For instance, a ``plus'' symbol typically indicates ``plus'' in calculator interfaces, but it means ``create a new task'' in a ToDo interface. This class-based annotation approach focuses on class labels rather than truly understanding the functionality of each element in the surrounding context, which may lead to a misunderstanding of the page layout.

To ensure the dataset is truly instructive rather than merely icon detection, we focus on describing screen status and element functionalities. Consider the above mentioned example: instead of providing a bounding box for the icon and labeling it as ``ICON\_PLUS'', we present the actual functionality of the element within its context (e.g., ``Create a new task''). This strategy offers a clear, instructive, and detailed description of the interactive element, enhancing the dataset's applicability. Besides, the traditional categories of ``IMAGE'' and ``TEXT'' provides very little information of the actual element functionality. Our annotation will provide a much more detailed and informative functionality for those elements (e.g., ``View the detail page of the 'Jazz Classics' playlist'' for the red left bottom ``compound'' element in Figure~\ref{fig:amex-elements}). Further elaboration on the collection and processing is provided in Appendix~\ref{app:desc-detail}.

\subsection{Level III: Instructions with GUI-Action Chains}
\label{sec:instrtions_with_gui_operations}

Table~\ref{tab:dataset-comparison} illustrates that most instruction-based datasets provide an average number of operation steps below 7, which implies most of the instructions are simpler than tasks in real-world. For example, users often apply filters or sort by different attributes when searching, which causes more complex instructions and longer action chains. In order to address this situation, we collect a set of complex instructions with GUI-action chains on popular apps from Google play store.

We define our action space for stepwise GUI operations similarly to \textsc{AitW}: \{\texttt{TAP}, \texttt{SCROLL}, \texttt{TYPE}, \texttt{PRESS\_BACK}, \texttt{PRESS\_HOME}, \texttt{PRESS\_ENTER}, \texttt{TASK\_COMPLETE}, \texttt{TASK\_IMPOSSIBLE}\}. \texttt{TAP} actions are characterized by identical touch and lift coordinates, while \texttt{SCROLL} actions involve distinct touch and lift coordinates. \texttt{TYPE} actions are annotated with a \texttt{type\_text} attribute specifying the input text. The three \texttt{PRESS} actions correspond to system-level button presses (back, home, enter). \texttt{TASK\_COMPLETE} and \texttt{TASK\_IMPOSSIBLE} serve as terminal flags for instructions. Additionally, for instructions that involve information query (e.g., ``What is the lowest price of the men's belt?''), we associate the \texttt{TASK\_COMPLETE} action with a region of interest, which is defined as the bounding box of the area on the current screenshot, where the answer is expected to appear (see examples in Appendix~\ref{app:more-exampls}). This comprehensive action space allows users to fully simulate a wide range of typical use cases. Detailed collection process and instruction generation are in Appendix~\ref{app:chain-detail}.

\section{Experiments}
\label{sec:experiments}

AMEX serves as a supplementary dataset for other large-scale instruction-based datasets. To test the effectiveness of AMEX, we directly finetune SPHINX models~\cite{liu2024sphinxx} without any tricks and conduct evaluation on three other benchmark datasets. We choose ScreenSpot~\cite{cheng-etal-2024-seeclick} as the benchmark of our agent's element grounding ability and \textsc{AitW} and \textsc{AndroidControl} as the benchmarks of our agent's GUI-control ability.

\subsection{SPHINX Agent}
\label{sec:agent-model}

Our SPHINX agent, SphAgent for short, is initialized from SPHINX~\cite{liu2024sphinxx} and it is finetuned for the GUI-control tasks and element grounding tasks. For both tasks, we train the model in a pure vision-based principle, which means the model understands the environment by only screenshots, without any other supplementary information such as accessibility trees or view hierarchy, due to the fact that many apps don't support or only provide very little accessibility tree or view hierarchy element information. We consider vision-based agents are more applicable in more various use cases.

\begin{table}
    \caption{ScreenSpot mobile subset experiment results.}
    \label{tab:screenspot-exp}
    \centering
    \resizebox{0.8\linewidth}{!}{%
    \begin{tabular}{l C{1cm} C{1.2cm} }
    \toprule
        Model & Size &  Icon / Widget \\
        \midrule
        Fuyu & 8B &  1.3\% \\
        CogAgent & 18B &  24.0\% \\
        SeeClick & 9.6B & 52.0\% \\
        Qwen2-VL & 7B &  60.7\% \\
        GPT-4V w. OmniParser & - &  57\% \\
        \midrule
        \textbf{SphAgent} & 7B  & \textbf{72.6\%} \\
        \bottomrule
    \end{tabular}
    }
\end{table}

\subsection{ScreenSpot}

\begin{table*}
    \caption{Experiment results on \textsc{AndroidControl}. ``Click \& Long Press'' is specially computed to show the improvements on these actions. The best results are in bold for different subsets and levels.}
    \label{tab:android-control-exp}
    \centering
    \resizebox{\linewidth}{!}{%
    \begin{tabular}{l c C{2.8cm} C{1.4cm} c C{1.5cm} C{1cm} C{1cm} c C{1.8cm}}
        \toprule
        Agent & \# data & Training Data & Task\newline Level & IDD & Category\newline Unseen & App\newline Unseen & Task\newline Unseen & Overall & Click \&\newline Long Press\\
        \midrule \midrule
        \multirow{2}{*}{SphAgent} & \multirow{2}{*}{178K} & \multirow{2}{*}{\textsc{AndroidCtrl}} & High Lv. & 58.3 & 40.6 & 39.4 & 52.2 & 49.8 & 34.1 \\
        & & & Low Lv. & 79.8 & 71.0 & 71.1 & 83.5 & 75.8 & 53.0 \\
        \midrule
        \multirow{2}{*}{SphAgent} & \multirow{2}{*}{178K} & \textsc{70\% AndrCtrl} & High Lv. & 60.4 & 41.7 & 43.4 & 59.3 & 52.8 & 37.3 \\
        & & + 10\% AMEX & Low Lv. & 81.9 & 73.2 & 72.6 & 87.5 & 77.7 & 56.7 \\
        \midrule
        \multirow{2}{*}{SphAgent} & \multirow{2}{*}{712K} & \textsc{AndroidCtrl} & High Lv. & \textbf{70.5} & \textbf{51.6} & \textbf{55.6} & \textbf{70.2} & \textbf{61.7} & \textbf{50.8} \\
        & & + AMEX & Low Lv. & \textbf{88.9} & \textbf{81.8} & \textbf{76.9} & \textbf{92.6} & \textbf{84.2} & \textbf{67.0} \\
        \bottomrule
    \end{tabular}
    }
\end{table*}

ScreenSpot~\cite{cheng-etal-2024-seeclick} is a benchmark dataset which contains over 600 UI screenshots from mobile devices, desktops and websites. It is designed to evaluate the model's element grounding capability, where each human-annotated functionality corresponds to an interactive element. We train SPHINX agent on AMEX level 1 and level 2 data to evaluate the element grounding capability.

Table~\ref{tab:screenspot-exp} lists the evaluation results on the ScreenSpot mobile subset. The models~\cite{fuyu-8b,hong2024cogagent,cheng-etal-2024-seeclick,wang2024qwen2} in first four rows are LVLMs specially trained on GUI data. OmniParser~\cite{wan2024omniparser} is a combination of an icon detection model, an icon description model and an OCR module, which serves as a screenshot parser for other LVLMs such as GPT-4V. During evaluation, our SphAgent surpasses four LVLMs by a large margin on the ``Icon / Widget'' subset without any training tricks, proving that the functionality understanding can largely boost the performance of GUI grounding for agents. The above results and comparisons strongly prove the effectiveness and applicability of AMEX in GUI element grounding tasks.

\subsection{\textsc{AndroidControl}}

\textsc{AndroidControl}~\cite{li2024androidcontrol} is a large-scale benchmark dataset to evaluate the agent's GUI-control ability to complete tasks on general apps (see more detailed description of the dataset and splits in Appendix~\ref{app:android-control-explain}). We evaluate three SphAgents trained on (i) only \textsc{AndroidControl}, (ii) random 70\% \textsc{AndroidControl} data mixed with random 10\% AMEX level 1 and 2, (iii) a mix of \textsc{AndroidControl} and AMEX level 1 and 2. The (ii) experiment ensures the same number of data point as in (i) experiment. The action space of \textsc{AndroidControl} is different from AMEX level 3 and \textsc{AitW}, so we only use level 1 and level 2 data during the experiments (ii) and (iii).

Table~\ref{tab:android-control-exp} lists the evaluation results on \textsc{AndroidControl} test set. The evaluation results of experiment (i) and (iii) strongly support the effectiveness of AMEX level 1 and level 2. Adding full data of AMEX level 1 and level 2 leads to an average 10\% overall performance gain and strongly improves the ``click'' and ``long press'' actions by more than 14\%. Also, the comparative results from experiment (i) and (ii) shows that even using the same number of data points, replacing down-stream instructional task data with environment understanding data would also lift the performance of agents on both low-level and high-level at an average of 2.5\%. This performance gain indicates the effectiveness of our multi-level annotations.

\subsection{\textsc{AitW}}

\textsc{AitW} is another large-scale benchmark dataset to evaluate the agent's GUI-control ability to complete task goals. We evaluate three SphAgents trained on (i) only \textsc{AitW}, (ii) a mix of \textsc{AitW} and AMEX. Since the GUI-action chains in AMEX share the same action space with \textsc{AitW}, we utilize all three levels of AMEX in the mixed data.

Table~\ref{tab:aitw-exp} lists the evaluation results on \textsc{AitW} test set, which has five subsets for different types of tasks. SphAgent trained on the mixed data (\textsc{AitW} + AMEX) achieves the highest overall accuracy among all agents, and when compared to the SphAgent trained on only \textsc{AitW}, the accuracies in ``General'' and ``Single'' categories have a notable gain of around 5\% and the overall accuracy lifts by 2.5\%. That the gains in other categories are not remarkable is possibly due to the mis-aligned element-wise annotation and unregulated instruction operation annotation. In Appendix~\ref{app:aitw-explain} we further explore the dataset evaluation and explain the reason why the evaluation results might be misleading.

\begin{table*}
  \caption{Experiment results on \textsc{AitW}.
  }
  \label{tab:aitw-exp}
  \centering
  \resizebox{0.85\linewidth}{!}{%
  \begin{tabular}{lcccccccc}
    \toprule
    Agent  & Training Data & General & Install & G-Apps & Single & WebShopping & Overall\\
    \midrule

    SphAgent  & \textsc{AitW} & 68.2 & 80.5 & 73.3 & 85.4 & 74 & 76.28\\
    
    SphAgent  & \textsc{AitW} + AMEX & 73.1 & 80.6 & 73.4 & 90.8 & 75.8 & 78.72\\

    SphAgent  & AMEX & 51.5 & 55.6 & 57.1 & 61.9 & 55.1 & 56.2 \\

    \bottomrule
    
  \end{tabular}
  }

\end{table*}


\begin{table*}
  \caption{Experiment results on \textsc{AMEX}. The test apps are excluded during the training of SphAgent on AMEX.
  }
  \label{tab:amex-exp}
  \centering
  \resizebox{\textwidth}{!}{%
  \begin{tabular}{lccccccccccc}
    \toprule
    Agent & Training Data & Gmail & Booking & YT-Music & SHEIN & NBC & CM & ToDo & Signal & Yelp & Overall \\
    \midrule

    SphAgent & \textsc{AitW} & 32.1 & 45.9 & 46.1 & 35.1 & 48.3 & 61.1 & 55.9 & 43.3 & 42.9 & 45.6 \\

    SphAgent & AMEX & 63.7 & 67.4 & 76.1 & 70.0 & 68.5 & 62.7 & 78.6 & 68.2 & 68.9 & 69.3 \\ 
    
    \bottomrule
    
  \end{tabular}
}
\end{table*}

\subsection{Cross-Domain Experiments}

Since AMEX Level 3 shares the same action space only with \textsc{AitW}, we conducted two experiments: (i) training SphAgent on AMEX and testing it on \textsc{AitW}, and (ii) training SphAgent on \textsc{AitW} and testing it on AMEX. The results, as shown in the last row of Table~\ref{tab:aitw-exp}, indicate poor performance when the agent is trained only on AMEX and tested on \textsc{AitW}. Similarly, Table~\ref{tab:amex-exp} demonstrates poor performance when the agent is trained on \textsc{AitW} and tested on AMEX. This significant discrepancy in evaluation results is likely due to the domain gap between the two datasets. First, the styles of collecting GUI-action chains differ between the two datasets. For instance, annotators in \textsc{AitW} tend to directly  search for apps, whereas annotators in AMEX prefer navigating through the app library to locate apps. This inconsistency in interaction strategies poses a challenge for the agent, which evaluates tasks based on static frames rather than dynamically adapting to different usage patterns. Second, although both datasets share the same action space, they differ in terms of visual appearance and versioning. The apps in AMEX are updated to their latest versions, while those in \textsc{AitW} are based on older versions. These versioning differences lead to changes in the visual appearance of GUI elements (e.g., color schemes, icons, or button designs), which can negatively impact the agent’s ability to recognize and interact with them.

\section{Discussions}

\subsection{Limitations and Future Work}
\label{sec:limitation-future}

\paragraph{Multi-lingual} Most existing datasets are limited to English, with UGIF~\cite{gubbi-venkatesh-etal-2024-ugif} being a notable exception, as it includes instructions and screenshots in eight languages. The AMEX dataset contains a small number of screenshots in Chinese and Spanish, primarily due to strict registration and login requirements for Chinese apps and a lack of expertise in other languages. Future work should incorporate multi-lingual screenshots, functionalities, and instructions to create a more robust and comprehensive multi-lingual environment for GUI agents.

\paragraph{Online Evaluation} Due to the limitation that SPHINX is not deployable, we encountered issues trying online evaluation such as AndroidWorld~\cite{rawles2024androidworlddynamicbenchmarkingenvironment}. In future work, online dynamic evaluation is more accurate and simulate the real-world agent usage. 

\paragraph{Multi-Platform} Android is an open-source OS and the community provides well-developed tools for interactions and emulations. iOS is another widely used mobile device OS but it is restricted and hard to run test on. In the future, a dataset extended to other OS or platforms would provide more general information.

\subsection{Ethical Considerations}
\label{sec:ethical}

\begin{itemize}[leftmargin=5mm]
    \item The accounts registered and logged in are all for testing purposes, not including any personal information. The dataset doesn't contain any private or personal information.
    \item The dataset, if misused, could be exploited for undesirable purposes, such as anti-fraud mechanisms and anti-script verification codes (see Appendix~\ref{app:wrong-aitw-ethical}), potentially leading to harm.
    \item Annotators received remuneration in line with local wage standards for their annotation works.
\end{itemize}

\section{Conclusion}

As AI agents become more prevalent, mobile GUI agents are emerging as a research hotspot. To address the lack of fundamental understanding of GUI elements in existing datasets, we present the Android Multi-annotation EXpo (AMEX) dataset, which includes three levels of annotation to provide a more instructive and detailed understanding of UI screens and elements. Additionally, we introduce a state-of-the-art SPHINX Agent, which can serve as a baseline model for future research.

\bibliography{custom}

\appendix

\section{Appendix / supplemental material}

\subsection{Pipeline Details}
\label{app:collection-details}

The raw data is collected using Android emulators, specifically Android Virtual Device and Genymotion\footnote{https://www.genymotion.com/}. We developed tools to autonomously perform emulator operations and record human actions. These tools leverage Appium\footnote{https://appium.io/docs/en/latest/}, an open-source, cross-platform test automation tool for Android, iOS, and web apps, under Apache license. Appium collects device screenshots during operations, each of which also corresponds to an XML file recording the basic screen layout with element attributes, such as bounding boxes and in-app descriptions.

\subsubsection{Collection details}

As mentioned in Section~\ref{sec:data-collection}, the collection of screenshots and their XML data utilizes two methods. Here, we provide a more detailed pipeline for executing an autonomous script to perform operations on an Android emulator. The script is designed to traverse an app in an unconstrained manner and collect data. It performs actions (see Appendix~\ref{app:script-detail}) at regular time intervals to allow each page to fully load. Then the script captures the current screenshot along with the corresponding XML file. Upon the raw data (yellow part in Figure~\ref{fig:pipeline}), we extract bounding boxes for interactive elements from the XML associated with each screenshot. Then annotators manually inspect each screenshot, identifying and selecting only the elements that are actually visible within the interface (see Appendix~\ref{app:element-filter} for detailed reasons). 

\subsubsection{Autonomous script details}
\label{app:script-detail}

The autonomous script controls the emulator using three actions: \texttt{TAP}, \texttt{SCROLL}, \texttt{TYPE}.

\begin{itemize}[leftmargin=5mm]
    \item For the \texttt{TAP} action, we employ two algorithms. The first randomly selects a clickable element on the current screen, while the second computes the index of a clickable element using a formula to ensure the elements chosen are likely unique. We apply one of these algorithms randomly for different executions.
    \item For the \texttt{SCROLL} action, we classify whether an area is vertically or horizontally scrollable by setting a width-height ratio threshold, $\mathcal{R}$. If the element's ratio exceeds $\mathcal{R}$, it is considered horizontally scrollable; otherwise, it is vertically scrollable. We then randomly select a scrollable element on the current screen and perform a scroll action based on its type.
    \item For the \texttt{TYPE} action, we pre-define a list of phrases relevant to the category of apps being tested. For example, [``Women's dress'', ``Nike sneakers'', ...] for clothing shopping apps. These phrases are primarily used in search scenarios. 
\end{itemize}

\subsubsection{GUI clickable element filtering}
\label{app:element-filter}

\begin{figure*}
    \centering
    \begin{subfigure}[b]{0.25\linewidth}
        \centering
        \includegraphics[width=1.0\linewidth]{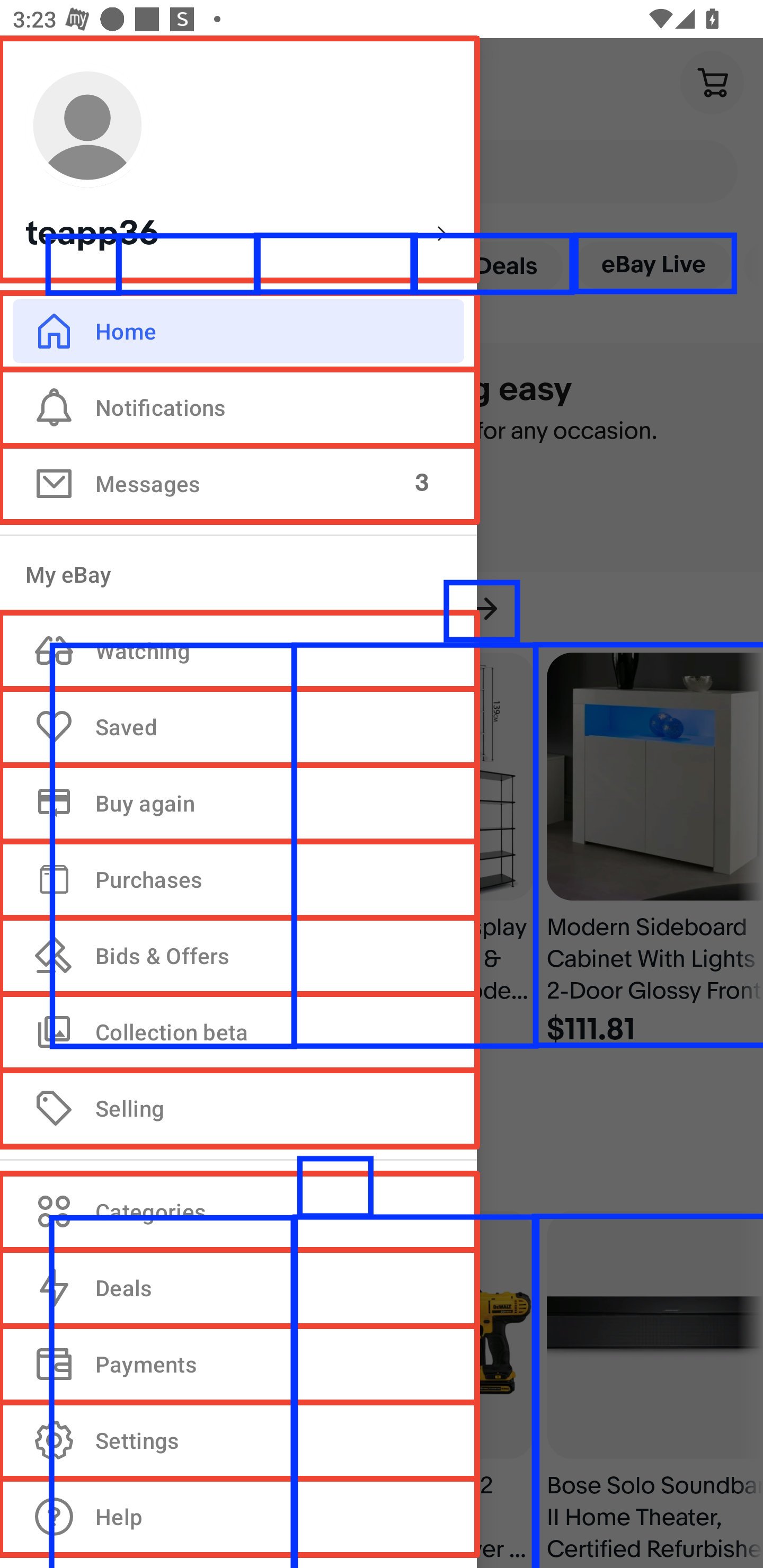}
        \caption{}
        \label{fig:before-filter}
    \end{subfigure}
    \hspace{3em}
    \begin{subfigure}[b]{0.25\linewidth}
        \centering
        \includegraphics[width=1.0\linewidth]{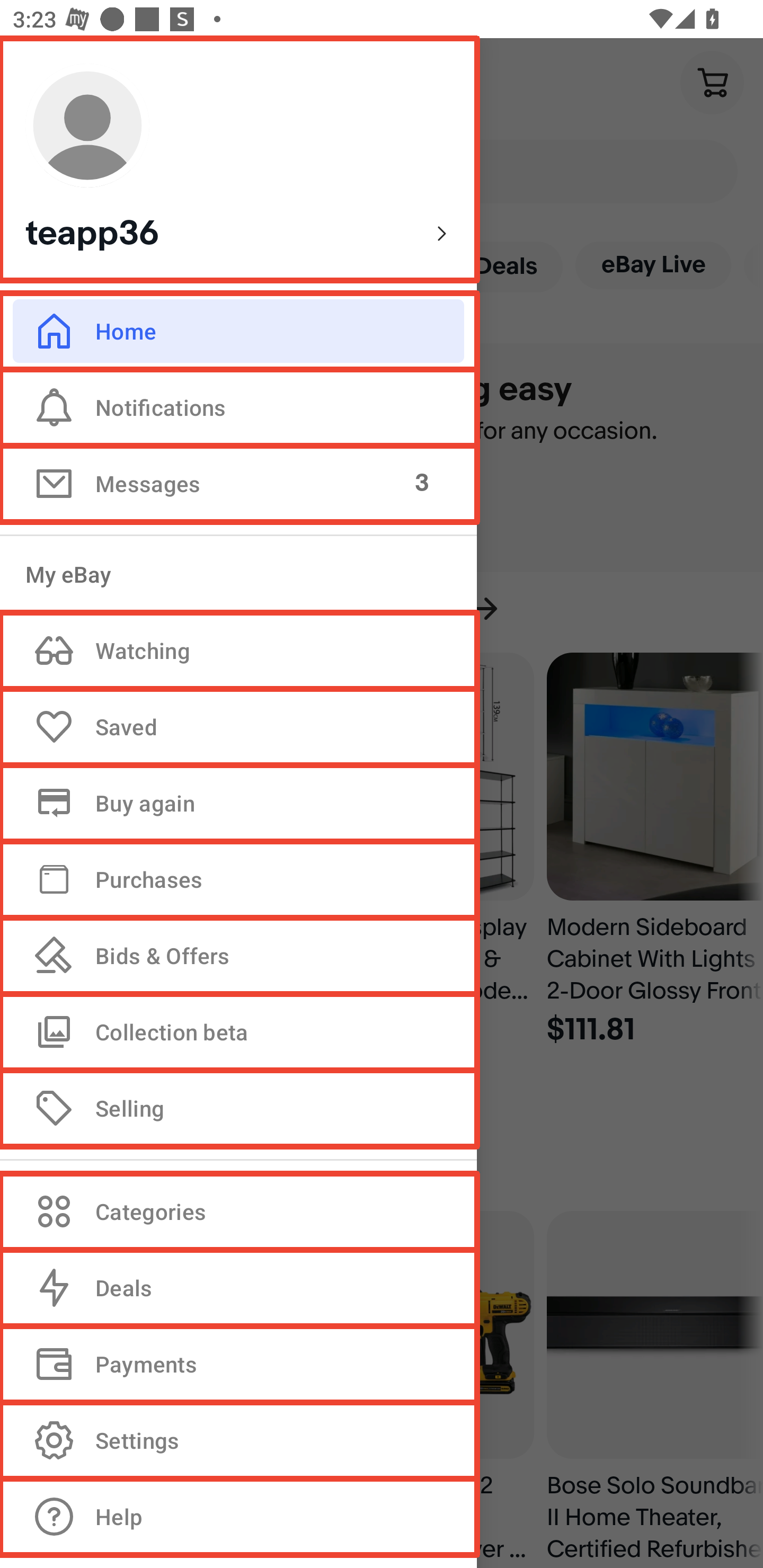}
        \caption{}
        \label{fig:after-filter}
    \end{subfigure}
    \caption{Demonstration of human annotator filtering. (a) before filtering. (b) after filtering.}
    \label{fig:filter}
\end{figure*}

The raw XML information sometimes contains the elements that are covered by other elements or layers. Figure~\ref{fig:filter} illustrates the same image before and after filtering. Blue boxes in Figure~\ref{fig:before-filter} are those elements under the current active layer and they still show up from XML parsing. Thus we need human annotators to filter out these blocked elements to get Figure~\ref{fig:after-filter}.

\subsubsection{GUI screen and element functionality description collection details}
\label{app:desc-detail}

We adopt GPT-4o and Gemini 1.5 Pro to describe screens and interpret element functionalities separately. Given the compact and dense nature of GUI elements, we apply Set of Mark (SoM) techniques~\cite{yang2023set} to boost GPT and Gemini's capability for visual localization. The coordinates of these elements are obtained and cleaned in the preceding phases (Section~\ref{sec:gui_interactive_element_grounding}). Additionally, recognizing that some elements possess abstract functionalities that are challenging to discern, we supplement the prompts for GPT and Gemini with in-app descriptions extracted from XMLs to enhance their comprehension of the screen. After the generation, we further utilize GPT-4o to compare the functionalities of the same element generated by two models and keep them if two functionalities have the same meaning. Finally, human annotators verify the quality and the functionality accuracy rate is above 97\%. 

\begin{figure*}
    \centering
    \begin{subfigure}[b]{0.48\linewidth}
        \centering
        \includegraphics[width=0.48\textwidth]{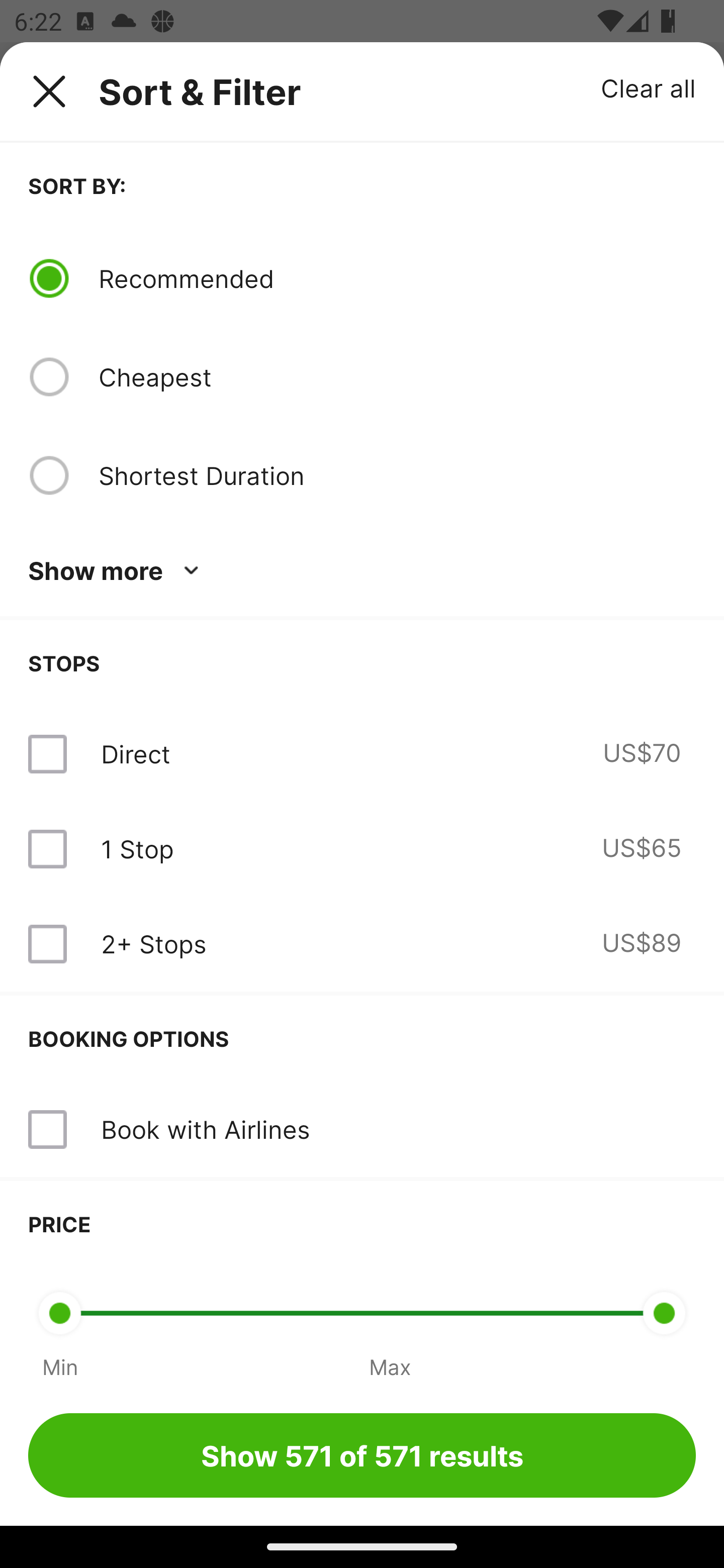}
        \includegraphics[width=0.48\textwidth]{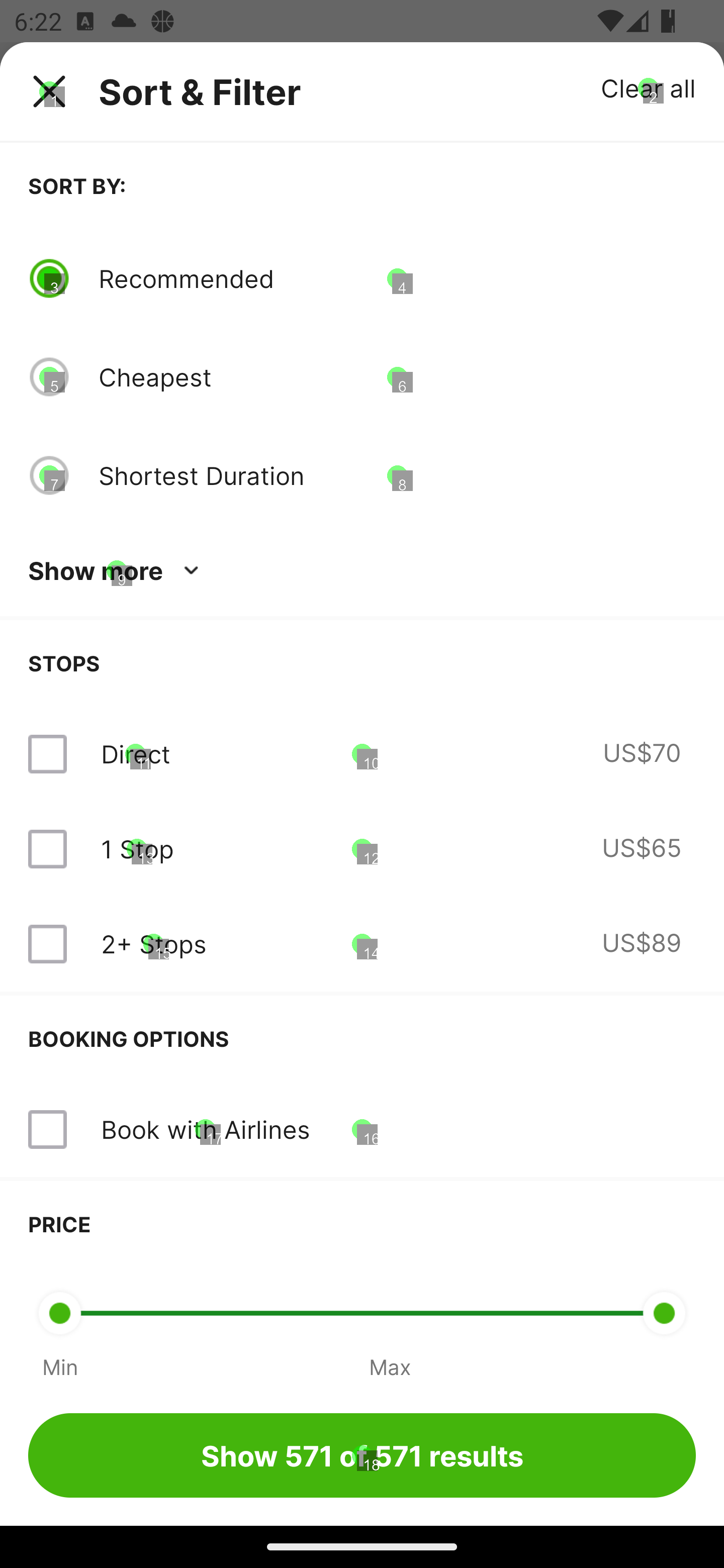}
        \caption{}
    \end{subfigure}
    \hfill
    \begin{subfigure}[b]{0.48\linewidth}
        \centering
        \includegraphics[width=0.48\textwidth]{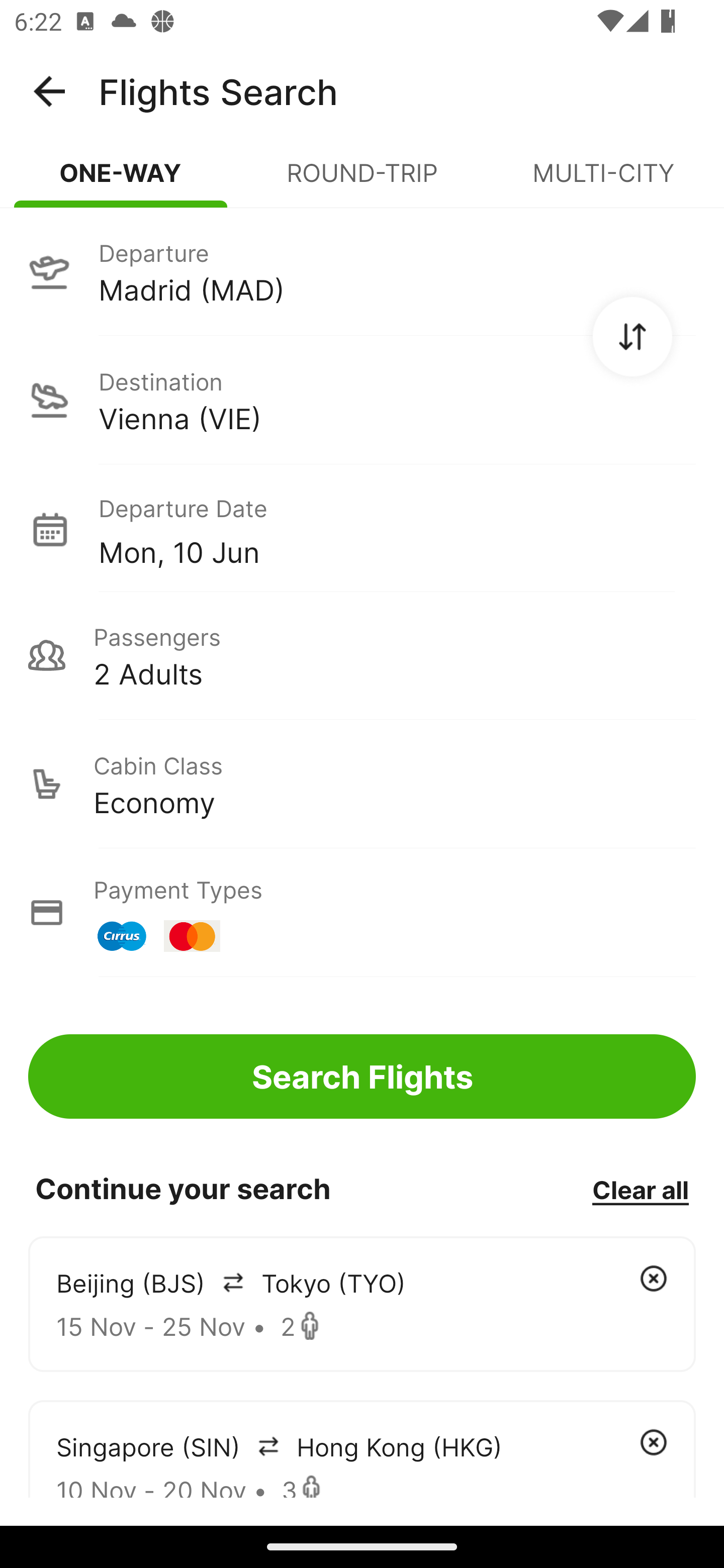}
        \includegraphics[width=0.48\textwidth]{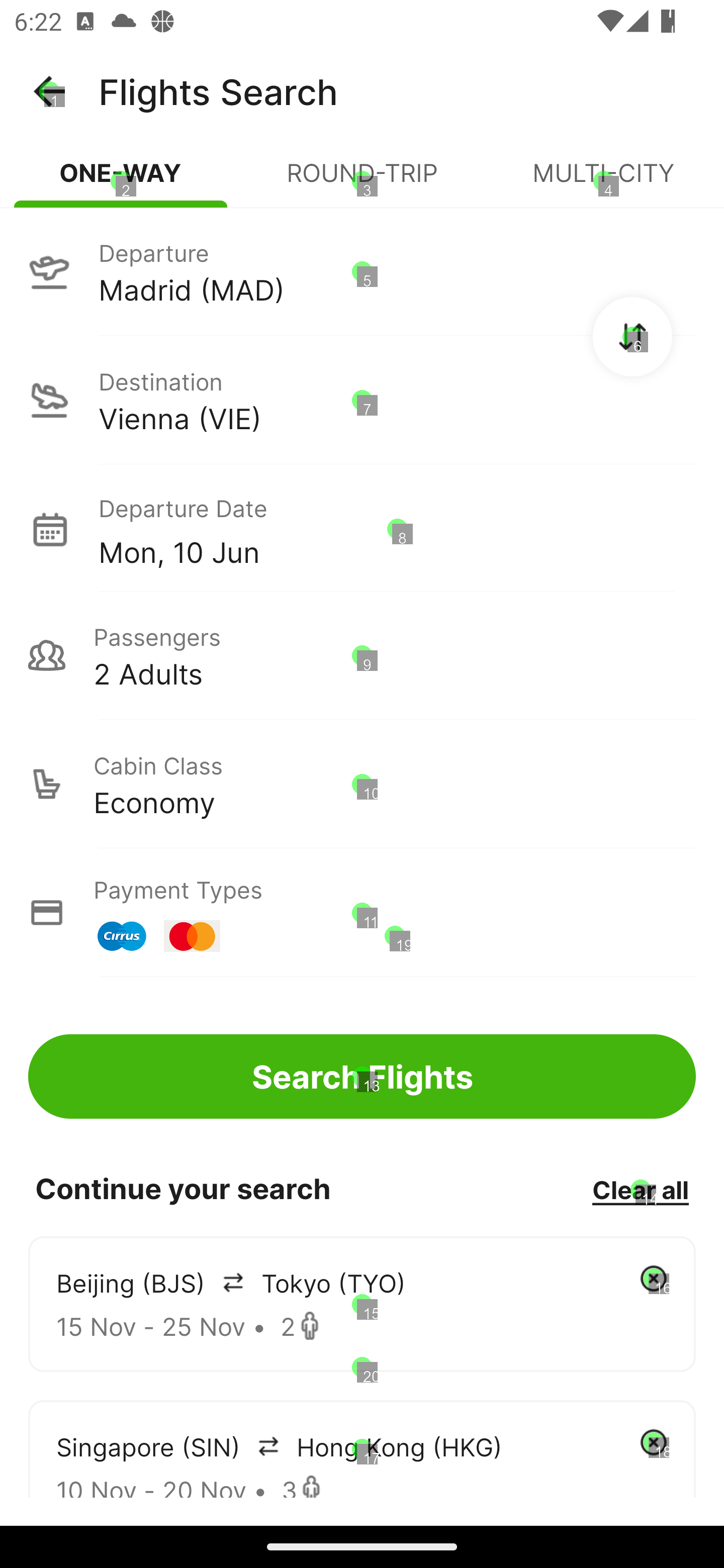}
        \caption{}
    \end{subfigure}
    \caption{Demonstrations of SoM technique on screenshots.}
    \label{fig:som}
\end{figure*}

We apply the Set-of-Mark (SoM) technique when using GPT and Gemini as the description generator. The SoM technique is a visual prompting method designed to enhance the visual grounding capabilities of large multimodal models (LMMs), such as GPT-4V, by overlaying visual marks on image regions. This involves partitioning an image into semantically meaningful regions and adding distinct marks (e.g., alphanumeric characters, masks, or boxes) to these regions. It demonstrates significant improvements in precision and accuracy over traditional prompting methods and other state-of-the-art models. Figure~\ref{fig:som} shows the screenshot with SoM technique.

The most recurring mistake is the mis-interpretation of the content description. For instance, some ``compound'' elements, which may contain several text, rating stars, image, etc., have couple of content descriptions. Those mixed content descriptions may cause the misunderstanding of the real functionality of the compound element. In IMDB app, a row of movie entry sometimes would consist of a poster, a text name, ratings of the movie, and the price of renting. The desired functionality of the entry could be "view the detail page of the movie xxx", while GPT might generate "view the price for renting the movie xxx", which is not exactly what we want. However, we use two LLMs to double check their generations on the same element functionality and we only keep the elements with the same functionalities generated by two LLMs.

The prompt is in the following Listing~\ref{lst:gpt_prompt_for_grounding} format.

\begin{lstlisting}[
float=htb,
language=bash,
floatplacement=htbp,
frame=TRBL,
frameround=tftf,
belowskip=-1\baselineskip,
basicstyle=\ttfamily\scriptsize,
breakatwhitespace=true,
breaklines=true,
captionpos=b,
columns=flexible,
keepspaces=true,
tabsize=2,
showspaces=false,
showstringspaces=false,
showtabs=false,
label={lst:gpt_prompt_for_grounding},
caption=An example prompt for guiding GPT4o to generate the element functionalities for the given Screenshot. ,
abovecaptionskip=0pt,
belowcaptionskip=0pt]
<@ Based on the screenshot of an Android mobile phone from the \textcolor{blue}{APP\_NAME}, please follow the instructions:  @>

<@ \textbf{1. Understand the Page Content:} @>
<@ - Analyze the overall content of the page. @>
<@ - Provide a brief summary of the page content in 1-2 sentences. @>
<@ \textbf{2. Explain Highlighted Areas:} @>
<@ - Each highlighted area is either clickable or scrollable. @>
<@ - Treat each highlighted area as a unique and separate entity, using identifiers such as <Region 1> etc. @>
<@ - If the highlighted area is a general icon, provide its type first in the format ICON\_Magnifying\_Glass. @>
<@ - If the highlighted area is more complex, provide a brief description in the format Element('a poster of  @>
<@ the movie named <La La Land>'). @>
<@ - Explain the purpose or functionality of each highlighted area. In other words, what result will happen @>
<@  or what's the user's intention when the marked <@ area is clicked or scrolled? @> 
<@ - Some functionality may require an overall analysis, and try to give the functionality specifically and @> 
<@ related to the current screenshot. @>
<@ \textbf{3. Additional Information for Highlighted Areas:} @> 
<@ There are total {NUMBER} elements to annotate. @> 
<@    \textcolor{blue}{MarkerInformation} @>
<@ \textbf{4. Output Format:} @> 
<@The output should be in JSON format as follows: @>
<@ \{ @> 
<@    "overall\_page\_content": "1-2 sentences summarizing the page content", @>
<@   "Region 1": "ICON\_XXX\_XXX <functionality>: xxxxx", @>
<@    "Region 2": "Element('a poster of a movie xxxx')  @>
 <@               <functionality>: click to see the details and forward the purchase page of the movie xx", @>
<@ \}  @>

\end{lstlisting}

\subsubsection{GUI-Action chain collection details}
\label{app:chain-detail}

The instruction generation process comprises three phases. Initially, human annotators create 5-10 complex instructions for each target app, considering its specific functions and capabilities. These initial instructions, combined with relevant app metadata collected online, serve as input for GPT. It then generates a larger set of 80-100 instructions that exhibit similar structure and intent to the human-provided examples. However, due to GPT's limitations on understanding real-world constraints, human filtering is adopted to modify or remove any unreasonable or impractical instructions. For example, the human-proposed instruction ``\textit{Open Booking.com, search for stays in Washington DC from 2024 June 11 to June 15.}'' would generate a similar but impractical ``\textit{Open Booking.com, search for stays in New York from 2024 May 3 to May 7}'', which needs modification since the dates are not selectable when the task is generated.

Since the meta information online doesn't reflect all functions of an app and sometimes may include misleading information, the instruction generated by GPT may have two types of impracticabilities: (i) the task is unrelated to the app activity (ii) some of the steps in the instruction are impractical. For instance, an app ``SEPHORA'' only sells beauty related products, but GPT may generate tasks that searches for electronics. Another example would be about a ticket booking app ``SeekGeek'', which doesn't provide a sort option ``sort by price from high to low''. While human created task would include ``sort by price from low to high'', GPT would generate something with``sort by price from high to low''. This step is impractical so we would filter it out.

Human annotators are each assigned a random selection of apps and their associated instructions, which they are asked to complete in a natural manner. In contrast to the \textsc{AitW} dataset, our collection methodology allows annotators to make errors and take incorrect steps, leading to a greater prevalence of \texttt{PRESS\_BACK} actions. This is motivated by our observation that agents trained on existing datasets exhibit difficulty navigating back to previous pages due to insufficient experience with the \texttt{PRESS\_BACK} action. Additionally, after completing an information query task, annotators are asked to manually mark the region of interest on the screenshot using our annotation tool. 

\subsubsection{Collection resources details}
\label{app:resource}

\begin{itemize}[leftmargin=5mm]
    \item GUI interactive element grounding takes approximately 3000 human-hour to filter bounding boxes described in Appendix~\ref{app:element-filter}.
    \item GUI screen and element descriptions use GPT-4o and Gemini API, which consumes about 800 dollars.
    \item Instructions with GUI-action chains take approximately 300 human-hour.
\end{itemize}

\subsection{Experiment details}\label{app:exp}

In our implementation, we utilize the internlm-7b variant of the SPHINX model, as detailed in~\cite{liu2024sphinxx}. The pre-trained checkpoint for this model was sourced from the official repository mentioned in~\cite{team2023internlm}. For image processing, the input images, each sized $1024 \times 1024$, are segmented into sub-images. Visual features from these sub-images are extracted using two distinct visual encoders: DINOv2~\cite{oquab2024dinov2} and ConvNext~\cite{woo2023convnext}. To ensure compatibility in feature dimensions across different modules, linear projection layers are employed to align the channel dimensions. Regarding the model's parameter settings, as outlined in Section~\ref{sec:agent-model}, we configure the history window size to four. Additionally, we introduce a special token, \texttt{<ICON>}, specifically designed to identify interactive elements within the interface, strengthening the model's interpretability and responsiveness to user interactions. The agent is trained on a cluster with 3 nodes, each with eight NVIDIA A100 (80GB) GPUs.  The fine-tuning was completed in four epochs. 

\subsubsection{\textsc{AndroidControl} details}
\label{app:android-control-explain}

\textsc{AndroidControl} splits the test set into four categories, i.e., in-domain data (IDD), category unseen, app unseen, and task unseen. Besides, the instructions are split into two levels. The high-level instructions are simply the goal of the task and the low-level instructions are the guides for each step during the task execution. Low-level tasks are easier for agents due to the fact that the agent only need to perform the action specified by the step-wise low-level instruction, while the high-level tasks require the agents to perform actions based on the overall instruction. For example, a step-wise low-level instruction is ``\textit{click the profile element at the bottom right of the screen.}'', compared with high-level instruction ``\textit{Change my profile name to Unknown111}''.

\subsubsection{\textsc{AitW} details}
\label{app:aitw-explain}

\begin{table*}
  \caption{Human evaluation results on 5\%-subsampled \textsc{AitW} test sub-set. ``MMR'' stands for ``Mis-Match Random'' subset.}
  \label{tab:aitw-human-anno}
  \centering
  \resizebox{\textwidth}{!}{%
  \begin{tabular}{lccccccc}
    \toprule
    Subset & Action Sources & General & Install & G-Apps & Single & WebShopping & Overall\\
    \midrule

    \multirow{2}{*}{MMR} & SphAgent vs. Human & 90.91 & 83.65 & 87.50 & 87.68 & 82.39 & 86.43 \\
    &\textsc{AitW} Anno. vs. Human & 88.18 & 75.00 & 82.69 & 84.78 & 75.57 & 81.25 \\

    \midrule

    \multirow{2}{*}{Random} & SphAgent vs. Human & 92.31 & 93.75 & 94.33 & 94.83 & 94.55 & 93.95 \\
    &\textsc{AitW} Anno. vs. Human & 94.23 & 90.34 & 93.62 & 93.97 & 95.00 & 93.43 \\
    
    \bottomrule
    
  \end{tabular}
  }
\end{table*}

\begin{figure*}
    \centering
    \begin{subfigure}{1.0\linewidth}
        \centering
        \includegraphics[width=0.9\textwidth]{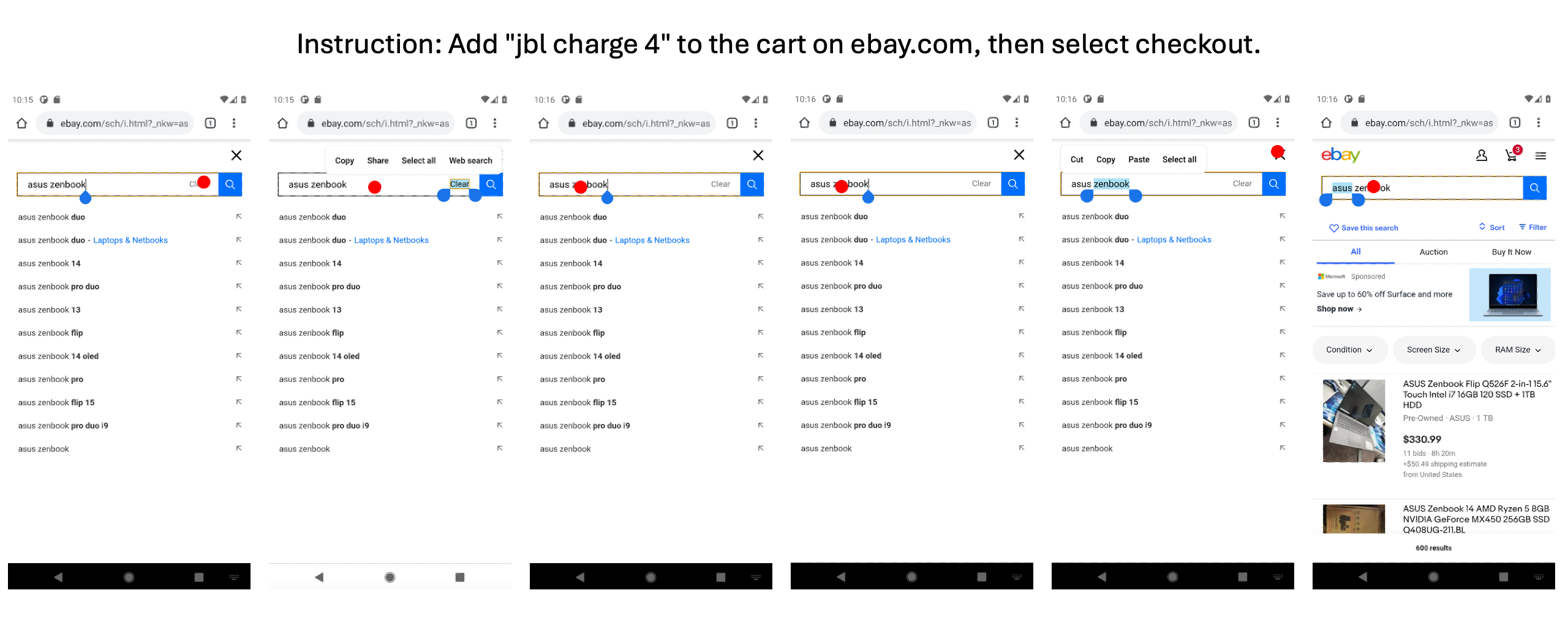}
        \caption{Repeating useless screenshots and actions.}
        \label{fig:repeat-aitw}
    \end{subfigure}
    \begin{subfigure}{1.0\linewidth}
        \centering
        \includegraphics[width=0.7\textwidth]{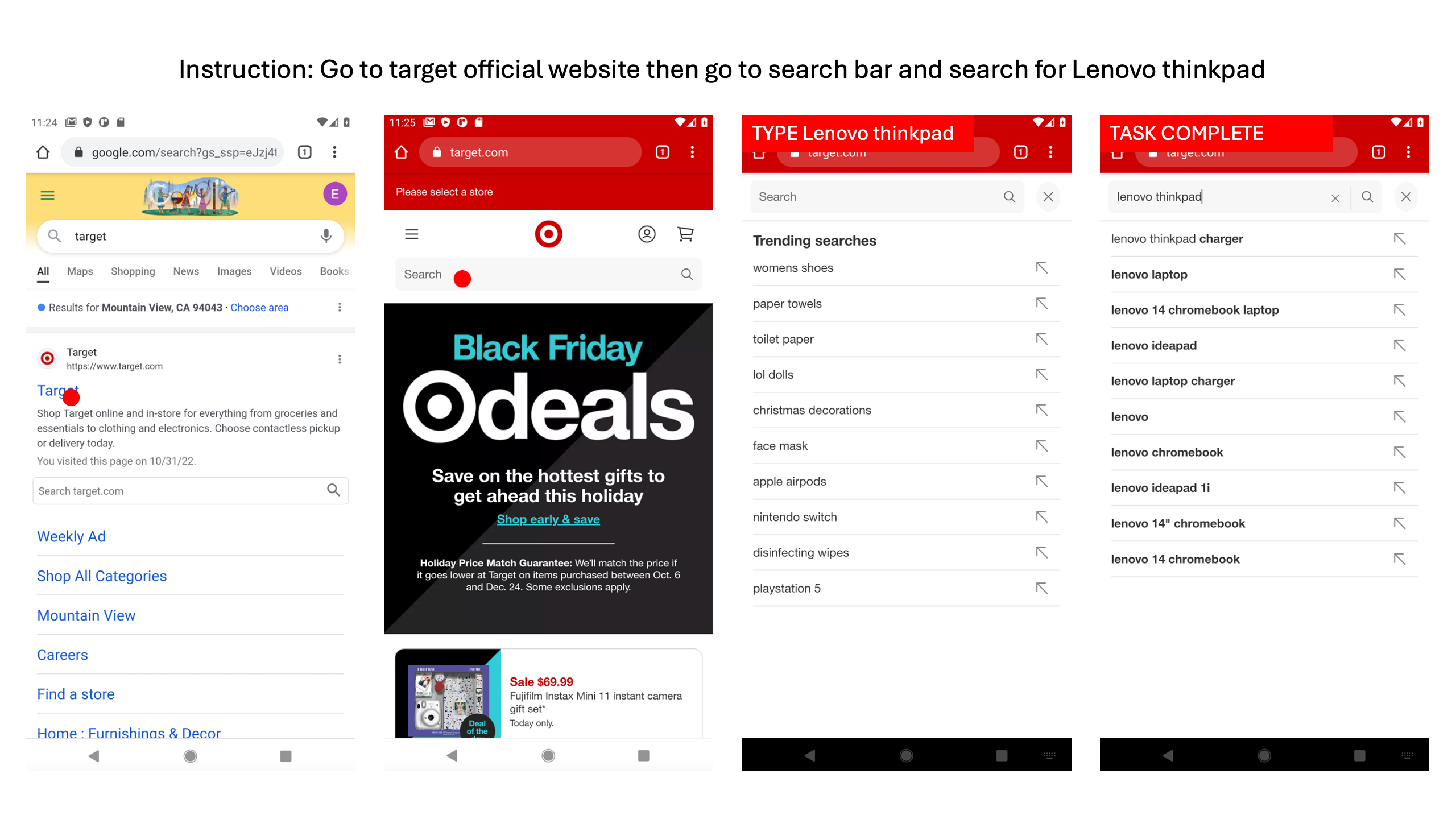}
        \caption{Wrong task complete.}
    \end{subfigure}
    \caption{Error cases in \textsc{AitW} test set. Red dots indicate the actions from the \textsc{AitW} annotations.}
    \label{fig:wrong-aitw}
\end{figure*}

\begin{figure*}
    \centering
    \includegraphics[width=0.9\linewidth]{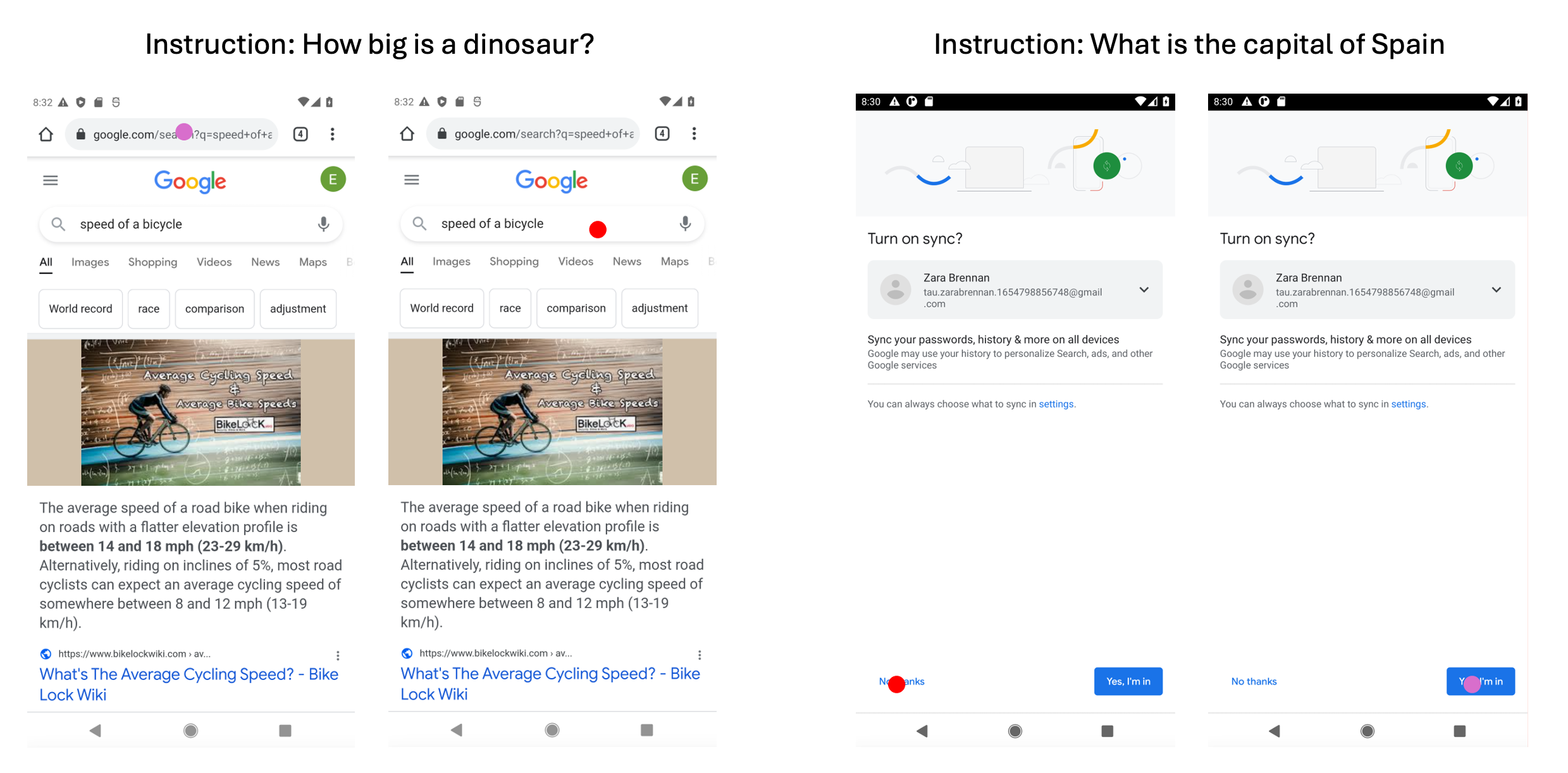}
    \caption{\textcolor{purple}{Purple} dots indicate the SphAgent inferred actions and \textcolor{red}{red} dots indicate the \textsc{AitW} annotations. Human annotators mark them both correct even though they are clicking different elements.}
    \label{fig:human-eval}
\end{figure*}

We observe that there are several types of error cases in the \textsc{AitW} test set (see Figure \ref{fig:wrong-aitw}). To assess the unreliability of the original \textsc{AitW} test set annotations, we hire human annotators to evaluate two subsets derived from the original test set. One subset is randomly chosen from episodes where SphAgent receives low accuracy (named \texttt{Mis-Match Random}), i.e., about 66\% compared to the overall 78.72\% shown in Table~\ref{tab:aitw-exp}, indicating a high mismatch between inference results and \textsc{AitW} annotations. The other subset is randomly selected from the remaining episodes (named \texttt{Random}). Annotators first filter out repeated and redundant screenshots (see Figure~\ref{fig:repeat-aitw}). For each remaining screenshot, the annotators then record the accuracies of SphAgent-inferred actions and the original \textsc{AitW} annotations versus our human evaluations. Figure~\ref{fig:human-eval} illustrates cases where annotators mark both the inferred action and the original annotation as correct, even though they interact with different elements. Table~\ref{tab:aitw-human-anno} presents the accuracy from human evaluation. The table indicates that the low SphAgent accuracy in the MMR subset are primarily due to unsatisfactory annotations (\textsc{AitW} Anno), which are used as ground truth during the evaluation. Both SphAgent and \textsc{AitW} Anno in the ``MMR'' subset have lower accuracies than those in the ``Random'' subset, highlighting that the tasks in the MMR subset are relatively more challenging. Furthermore, SphAgent achieves better human evaluation results than the original annotations, demonstrating its effectiveness and close alignment with human judgment. Comparing the results from Table~\ref{tab:aitw-exp} and Table~\ref{tab:aitw-human-anno}, the notable differences in overall scores (i.e., 78.72\%, 86.43\%, and 93.95\%) underscore the unreliability and misleading nature of the \textsc{AitW} test set annotations.

\subsection{More AMEX examples}
\label{app:more-exampls}

\begin{figure*}
    \centering
    \begin{subfigure}{0.95\linewidth}
        \centering
        \includegraphics[width=1.0\textwidth]{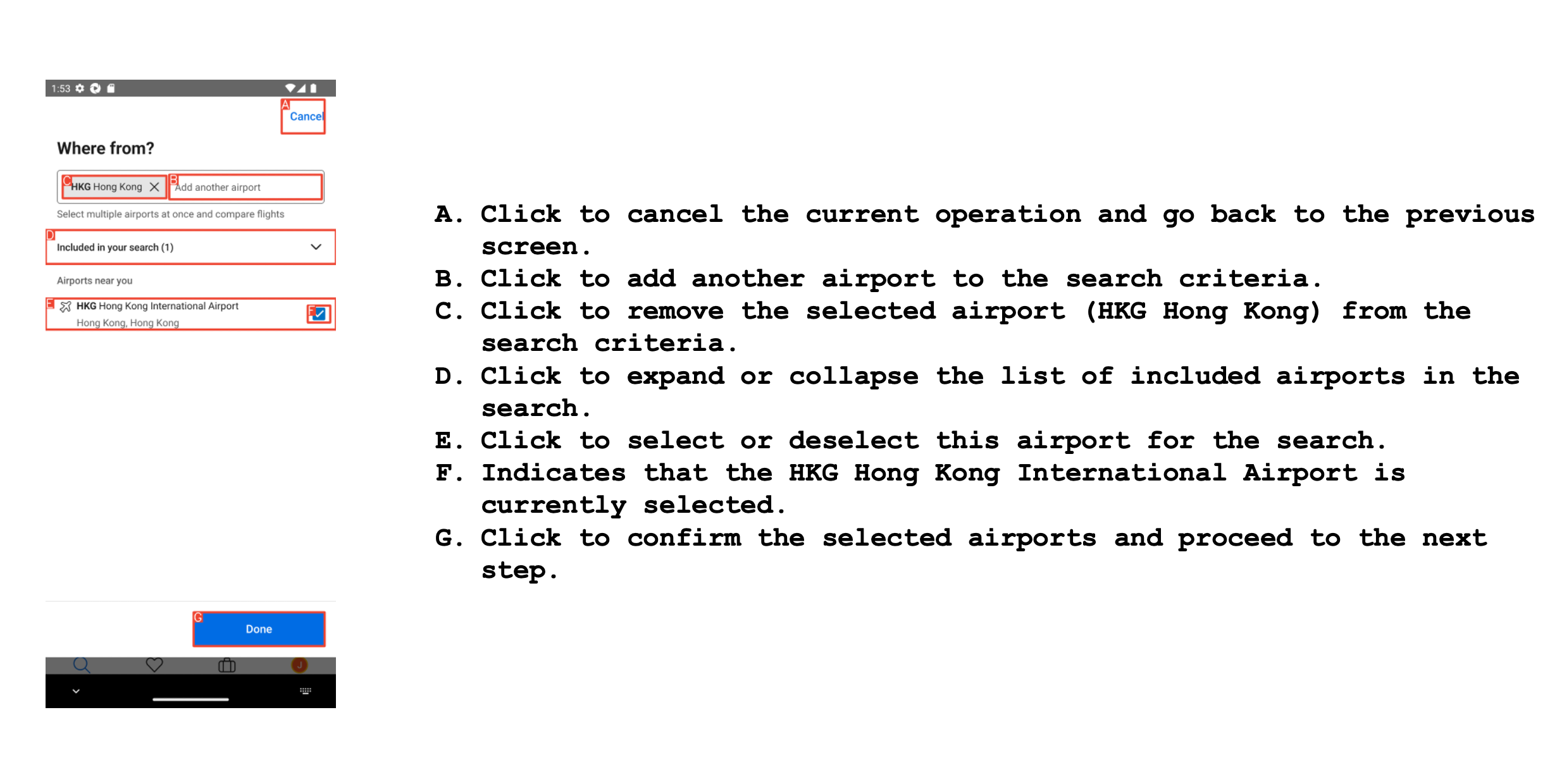}
    \end{subfigure}
    \begin{subfigure}{0.95\linewidth}
        \centering
        \includegraphics[width=1.0\textwidth]{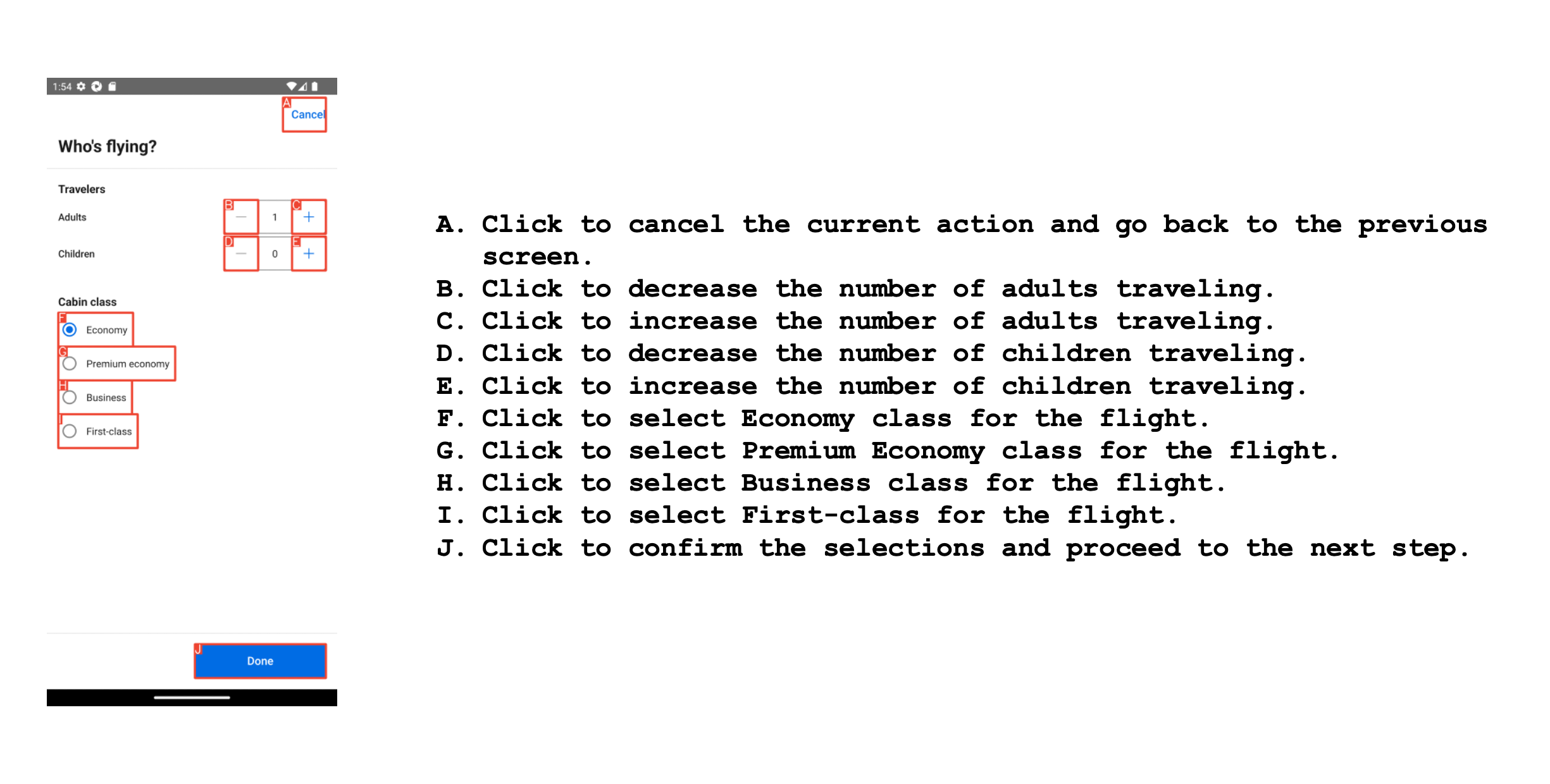}
    \end{subfigure}
    \begin{subfigure}{0.95\linewidth}
        \centering
        \includegraphics[width=1.0\textwidth]{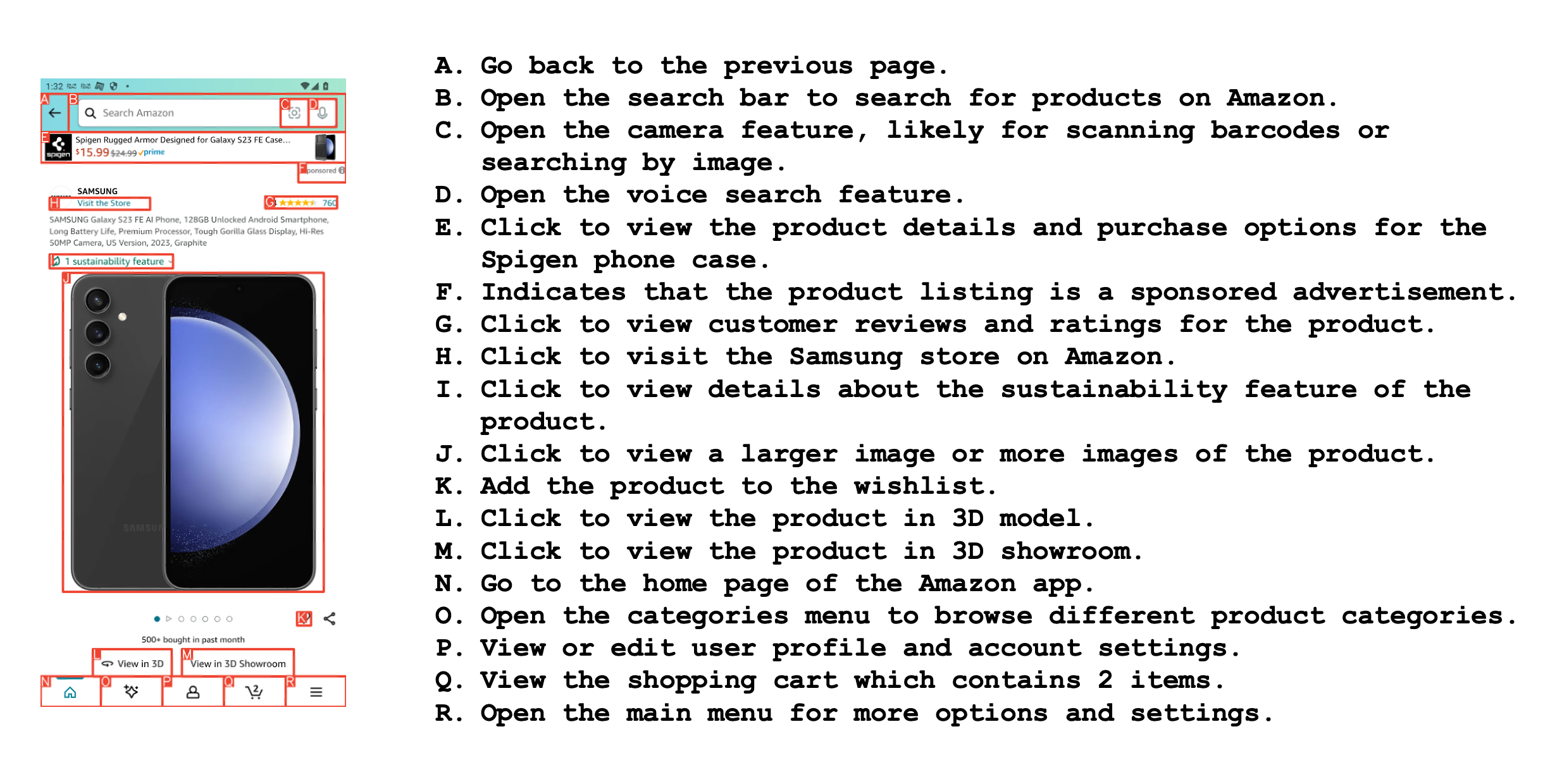}
    \end{subfigure}
    \caption{More examples}
    \label{fig:more-element}
\end{figure*}

\begin{figure*}
    \centering
    \begin{subfigure}{1.0\linewidth}
        \centering
        \includegraphics[width=1.0\textwidth]{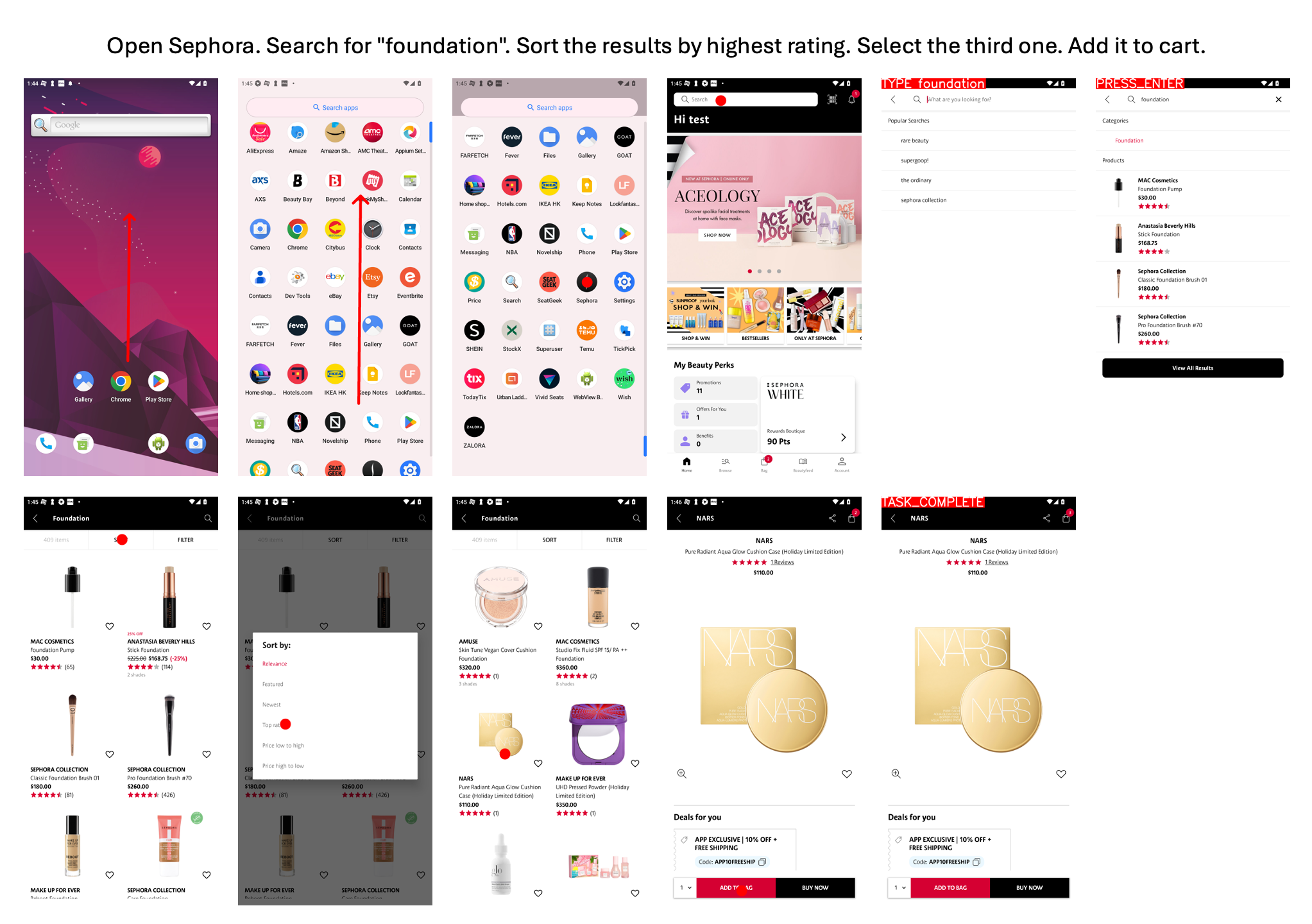}
    \end{subfigure}
    \begin{subfigure}{1.0\linewidth}
        \centering
        \includegraphics[width=1.0\textwidth]{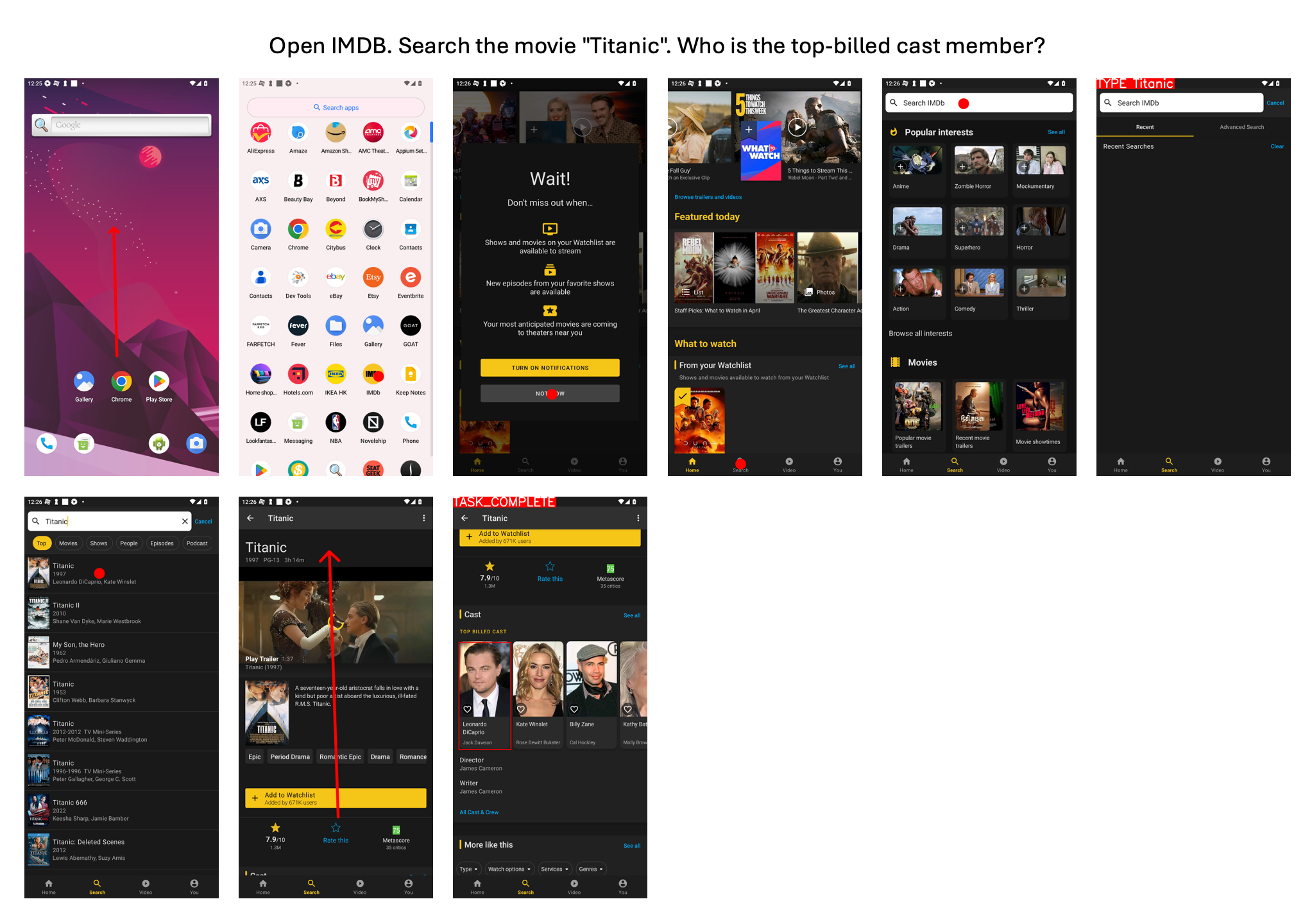}
    \end{subfigure}
    \caption{More examples}
    \label{fig:more-inst1}
\end{figure*}

See Figure~\ref{fig:more-element} for more examples of GUI interactive elements grounding and description. See Figure~\ref{fig:more-inst1} for more examples of instruction with GUI-action chains.

\subsection{Examples of Ethical Problems}
\label{app:wrong-aitw-ethical}

\begin{figure*}
    \centering
    \begin{subfigure}[b]{0.285\linewidth}
        \centering
        \includegraphics[width=1.0\textwidth]{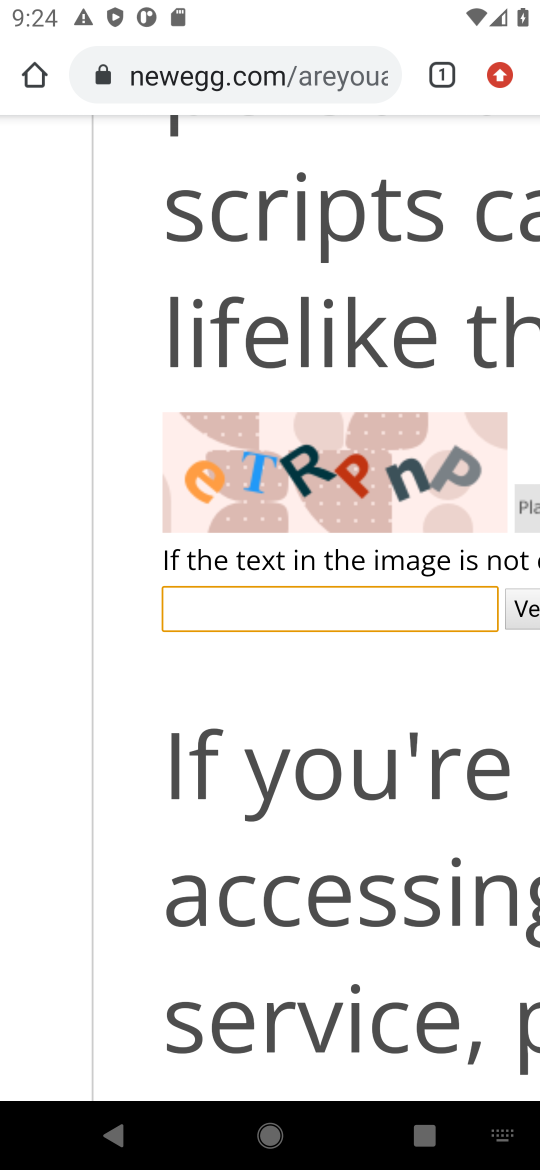}
    \end{subfigure}
    \hspace{1em}
    \begin{subfigure}[b]{0.3\linewidth}
        \centering
        \includegraphics[width=1.0\textwidth]{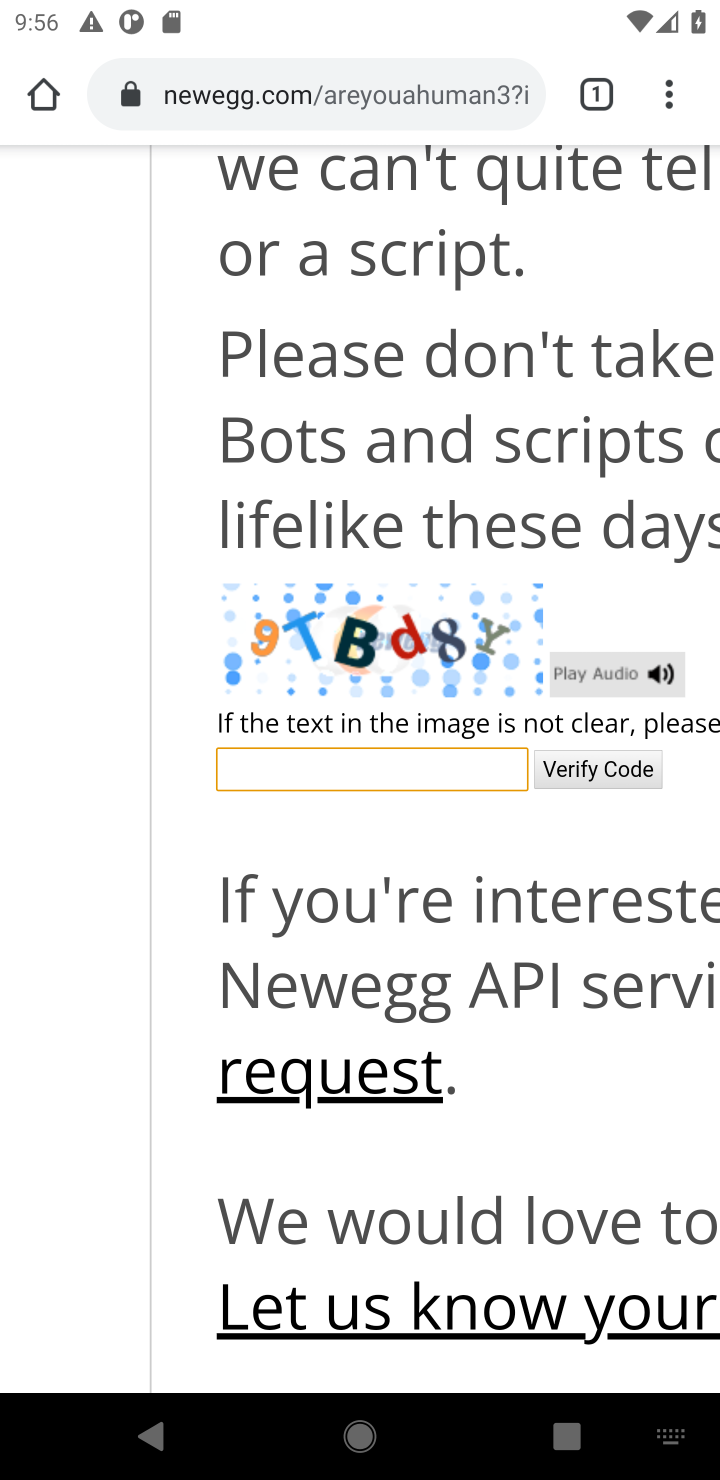}
    \end{subfigure}
    \hspace{1em}
    \begin{subfigure}[b]{0.3\linewidth}
        \centering
        \includegraphics[width=1.0\textwidth]{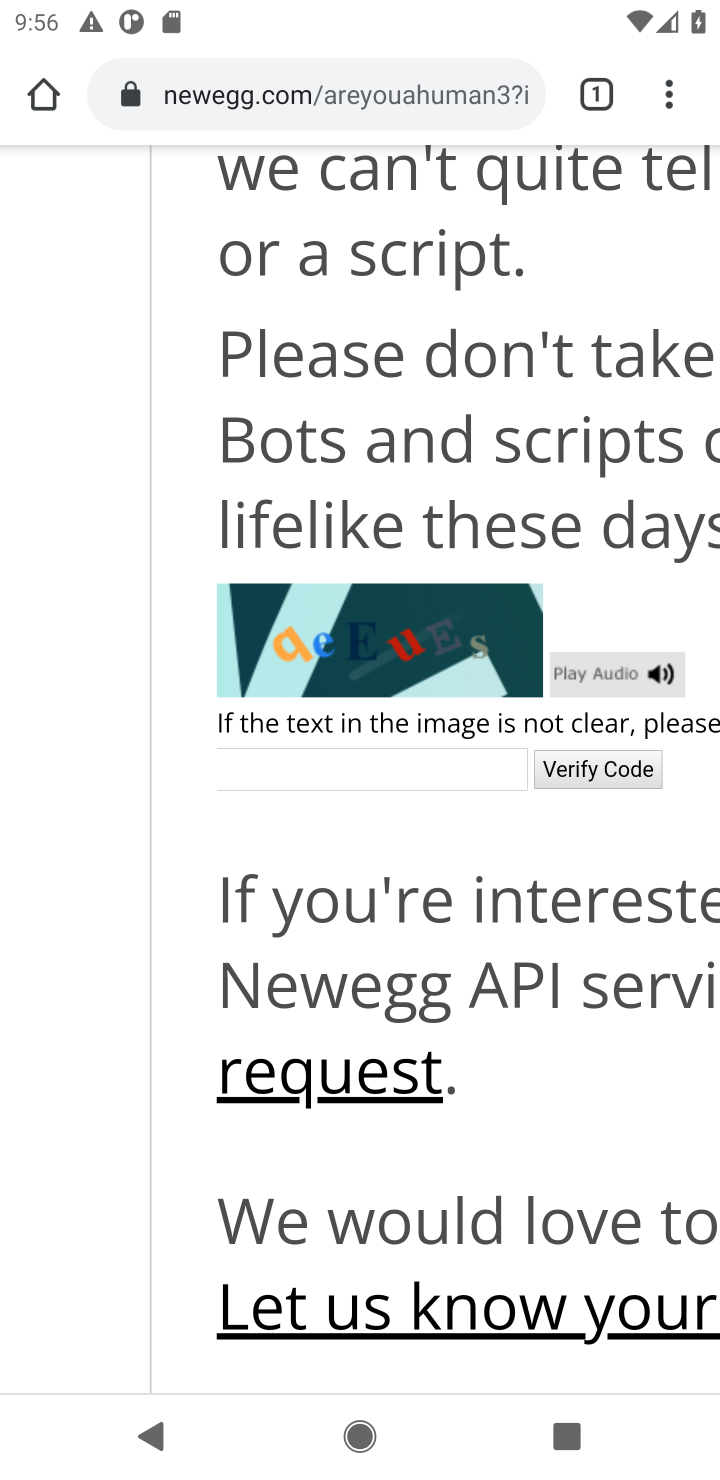}
    \end{subfigure}
    \caption{Demonstration of anti-script mechanism where agents can enter the verification codes correctly.}
    \label{fig:anti-script}
\end{figure*}

Figure~\ref{fig:anti-script} shows examples of anti-script mechanism where the agent can correctly enter the verification codes.

\end{document}